\documentclass[conference]{IEEEtran}
\IEEEoverridecommandlockouts
\usepackage{cite}
\usepackage{amsmath,amssymb,amsfonts}
\usepackage{algorithm}
\usepackage{algpseudocode}
\usepackage{etoolbox}
\usepackage{url}
\usepackage{graphicx}
\usepackage{textcomp}
\usepackage{xcolor}
\usepackage{multirow}
\usepackage{subcaption}
\usepackage{array}
\usepackage{longtable}
\def\BibTeX{{\rm B\kern-.05em{\sc i\kern-.025em b}\kern-.08em
    T\kern-.1667em\lower.7ex\hbox{E}\kern-.125emX}}
    
\newcommand{\Section}[1]{\vspace{+1ex}\section{#1}\vspace{+1ex}}

\title{AdaCache: A Disaggregated Cache System with Adaptive Block Size for Cloud Block Storage}

 \author{\IEEEauthorblockN{Qirui Yang}
 \IEEEauthorblockA{
 \textit{Samsung}\\
 qirui.y@samsung.com
 }\and
 \IEEEauthorblockN{Runyu Jin}
 \IEEEauthorblockA{
 \textit{Arizona State University}\\
 runyu.jin@asu.edu
 }\and
 \IEEEauthorblockN{Ni Fan, Devasena Inupakutika, Bridget Davis}
 \IEEEauthorblockA{
 \textit{Samsung} \\
 ni.fan, devasena.i, b.davis@samsung.com
 }\and
 \IEEEauthorblockN{Ming Zhao}
 \IEEEauthorblockA{
 \textit{Arizona State University}\\
 mingzhao@asu.edu
 \thanks{Partly supported by National Science Foundation 2126291, 1955593}
 }}
\begin{document}
\maketitle
\IEEEpubidadjcol

\begin{abstract}
NVMe SSD caching has demonstrated impressive capabilities in solving cloud block storage's I/O bottleneck and enhancing application performance in public, private, and hybrid cloud environments. However, traditional host-side caching solutions have several serious limitations. First, the cache cannot be shared across hosts, leading to low cache utilization. Second, the commonly-used fix-sized cache block allocation mechanism is unable to provide good cache performance with low memory overhead for diverse cloud workloads with vastly different I/O patterns. This paper presents AdaCache, a novel userspace disaggregated cache system that utilizes adaptive cache block allocation for cloud block storage. First, AdaCache proposes an innovative adaptive cache block allocation scheme that allocates cache blocks based on the request size to achieve both good cache performance and low memory overhead. Second, AdaCache proposes a group-based cache organization that stores cache blocks into groups to solve the fragmentation problem brought by variable-sized cache blocks. Third, AdaCache designs a two-level cache replacement policy that replaces cache blocks in both single blocks and groups to improve the hit ratio. Experimental results with real-world traces show that AdaCache can substantially improve I/O performance and reduce storage access caused by cache miss with a much lower memory usage compared to traditional fix-sized cache systems.
\end{abstract}


\Section{Introduction}
\label{sec:intro}
Block storage is widely used in public, private, and hybrid cloud environments because it is highly effective in providing fast, scalable, and reliable access to data~\cite{ebs, google, ibm, ceph, whatis}. Although cloud block storage is generally considered to be more I/O performant than other types of cloud storage such as object storage and file storage~\cite{ebsefss3}, it still falls short of the performance provided by directly attached NVMe SSD storage. To accelerate modern data-intensive applications such as Deep Learning (DL) training and big data processing~\cite{9925356}, NVMe SSD caching is employed to exploit workload locality for faster data accesses~\cite{osca, efficientssdcache, cloudcache, cachededup}. Typically, NVMe SSD cache devices are directly attached to each computing server which is usually multiple network hops away from storage servers~\cite{understandrackscale}. However, this host-side caching mechanism~\cite{cloudcache,cachededup} can lead to uneven cache utilization for two reasons. First, different computing servers run different cloud workloads can require varying degrees of cache resources. Second, a cache device is only used by the server where it is attached and cannot be shared or utilized across multiple computing servers.

By decoupling cache devices from computing servers, rack-scale cache disaggregation enables cache sharing through the pooling of cache resources within the same group of racks. Cache resources are managed and allocated as a whole which can lead to better cache utilization, scalability, and failure isolation. To achieve this, NVMe over Fabrics (NVMeoF)~\cite{nvmeof} can be employed to deliver high performance and scalability. NVMeoF defines a standard protocol for efficiently transporting the NVMe storage protocol over the network, which can scale out to large numbers of NVMe devices and extend the distance over which they can be accessed with low latency and high IOPS within a data center~\cite{nvmeofblog}. 

The fix-sized cache block management method commonly used in various cache system designs may not be the desirable solution for cloud workloads that are constantly changing. Using smaller cache blocks like 32KiB can achieve better I/O performance as it incurs smaller cache miss penalty~\cite{10.1145/633625.52433} compared to larger cache block sizes. However, its metadata overhead for managing the cache resource is higher, which causes larger memory footprint as the metadata usually needs to be cached in memory for performance. Conversely, using larger cache blocks such as 512KiB can improve the cache hit ratio~\cite{hennessy2011computer} by exploiting the spatial locality within the requests and reduce the memory overhead associated with metadata. However, this comes at the cost of larger cache miss penalty, which can significantly reduce I/O performance if the spatial locality is rare. 


In this paper, we aim to design a rack-scale disaggregated cache solution that provides good cache performance with low metadata overhead, regardless of the cloud workloads. We propose AdaCache, a rack-scale disaggregated cache system that employs variable-sized cache blocks to adapt to various cloud workloads. 
AdaCache allocates cache blocks of different sizes based on the I/O request size. For requests with large I/O sizes, large cache blocks are allocated to reduce the number of allocated cache blocks, thus improving I/O performance and reducing metadata memory overhead. For requests with smaller sizes, AdaCache assigns small cache blocks to avoid read/write amplification between the cache system and backend storage as well as cache pollution.

The contributions of this paper are as follows:
\begin{enumerate}
    \item The design and implementation of AdaCache, a practical rack-scale disaggregated cache system implemented using the SPDK framework~\cite{spdk} for cloud block storage. 
    \item The design of adaptive cache block allocation which incorporates three core ideas: efficient variable-sized cache block allocation algorithm, group-based cache organization, and two-level cache replacement.
    \item The comprehensive evaluation of AdaCache using publicly available real-world block I/O traces through both simulation and AdaCache prototype.
\end{enumerate}

According to the evaluation results, AdaCache has demonstrated significant improvements in I/O performance compared to traditional fix-sized cache. Specifically, it can improve read latency by 20\% and write latency by 9\% compared to 32KiB block-sized cache in trace replay. AdaCache is also capable of saving up to 74\% I/O traffic to cloud block storage and up to 63\% I/O traffic to the cache compared to 256KiB block-sized cache. Moreover, AdaCache has achieved up to 41\% memory savings compared to 32KiB block-sized cache. All of these improvements are accomplished with merely 2 microseconds of computation overhead at cache layer compared to a traditional fix-sized cache.

The rest of this paper is structured as follows. In Section~\ref{sec:background}, we introduce the design and implementation of disaggregated cache. In Section~\ref{sec:design}, we elaborate on the details of our AdaCache design. In Section~\ref{sec:eval}, we present our experimental method and the results. In Section~\ref{sec:related}, we discuss the related works and conclude in Section~\ref{sec:conclusion}.



\Section{Disaggregated Cache}
\label{sec:background}
\subsection{Rack-Scale Cache Disaggregation}

Cloud block storage has been widely adopted by today's public, private, and hybrid cloud infrastructure for primary data storage~\cite{ebs, ibm, google, ceph}.  With block storage, data is partitioned into fix-sized blocks and stored on the underlying storage medium. These blocks can be directly accessed by applications or through mounted file systems~\cite{yubometadata, hasfs}, allowing for quick modification of specific blocks to efficiently serve I/O requests.

NVMe SSDs are commonly used as a caching solution in large-scale cloud block storage systems to improve I/O performance~\cite{9091319}. Typically, caches are deployed on computing hosts to mitigate the high network latency to the storage clusters. However, cloud providers often encounter the challenge of load imbalance where some cache devices are more heavily used than others, leading to overloaded, under-loaded, or well-loaded cache devices on computing hosts~\cite{afzal2019load}. This results in unbalanced cache utilization and wasted cache resources.

\begin{figure}[t]
	\centering
	\includegraphics[width=0.9\columnwidth]{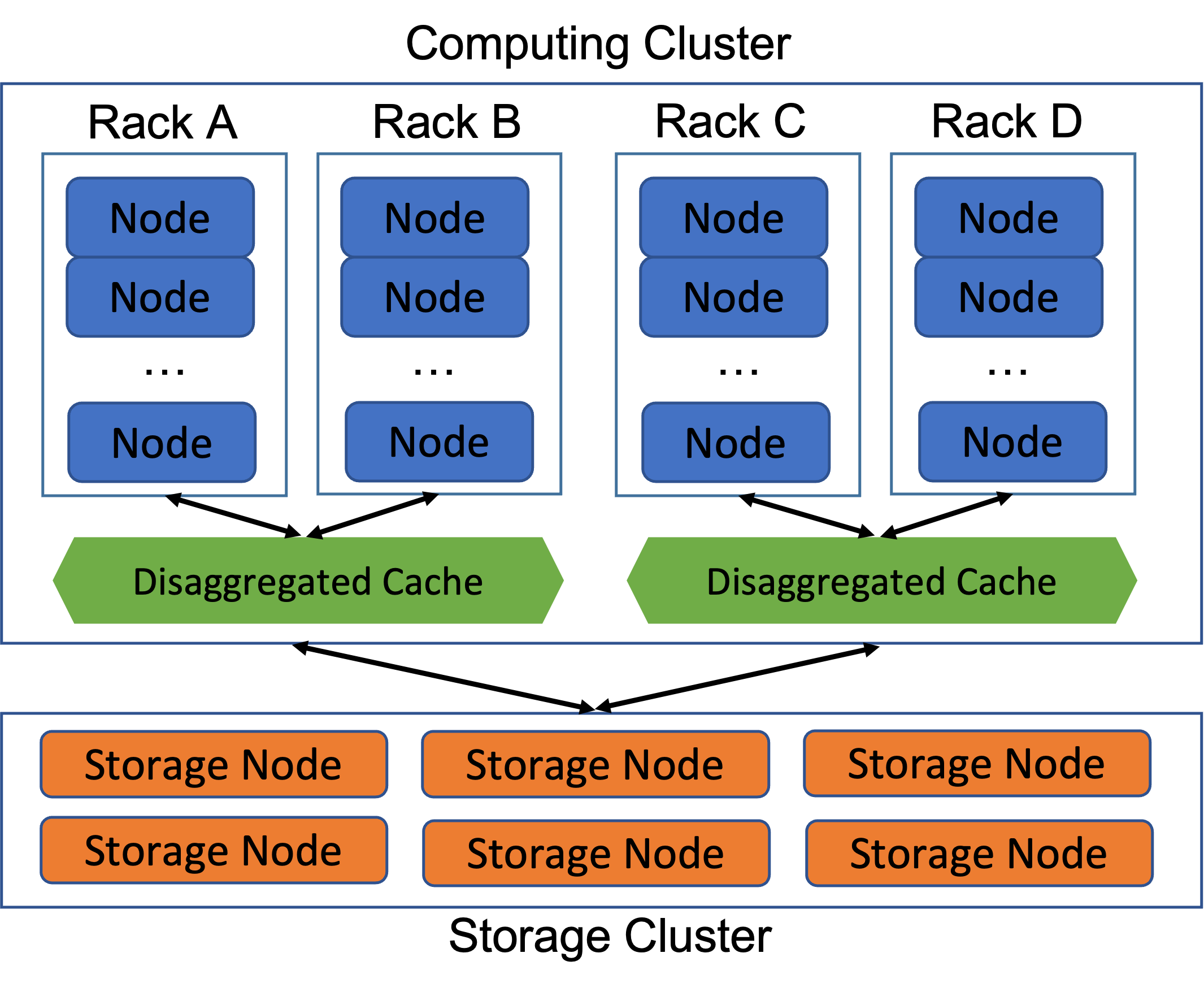}
	\caption{Rack-Scale Cache Disaggregation}
	\label{fig:rack}
    \vspace{-20pt}
\end{figure}

Cache disaggregation presents a solution to the aforementioned issues by disaggregating all the cache resources, enabling cache to be shared and managed as a whole.
It decouples SSD cache from the computing nodes and allows independent utilization of cache resources regardless of where an application is placed. In this sense, the cache resources are shared by all the applications and the cache load imbalance problem is addressed.
In cloud environments, this can be achieved at either cluster scale or rack scale. Cluster-scale cache disaggregation offers more pooled cache resources and consequently can result in better cache utilization compared to rack-scale. However, it suffers from higher network latency to access cache across the cluster which can negatively impact I/O performance. Additionally, it requires complicated software design and may inversely bring unacceptable software overhead and offset its benefit. Conversely, rack-scale cache disaggregation can provide superior cache resource utilization compared to the local cache and involve much lower network and software overhead compared to cluster-scale. As such, it provides an optimal trade-off between cache resource utilization and I/O performance. Figure~\ref{fig:rack} illustrates an example of rack-scale cache disaggregation. 

Rack-scale cache disaggregation enables cache devices within the same group of racks to share a cache server, providing computing servers of the same rack group with a pool of shared cache resources. The fast data transfer between computing nodes and the cache server can be achieved with the adoption of NVMe over Fabrics (NVMeoF)~\cite{nvmeof} technology, which is a protocol designed to provide storage to computing servers through the network using the NVMe protocol. It adds less than 10 microseconds of additional latency to locally attached NVMe devices~\cite{nvmeof10us}, making it an ideal choice for connecting the cache pool to the computing nodes. 
According to a recent performance report~\cite{nvmeofperformance}, NVMeoF using RDMA~\cite{nvmeof} has demonstrated impressive speed, achieving more than 11M 4K IOPS with an average latency of 231 microseconds using 100 Gbps NICs. As network bandwidth continues to double every few years, this performance is expected to improve even further. With such high performance, a single cache server can effectively serve thousands of concurrent NVMeoF connections. Furthermore, a single cache server can provide large storage capacities. For example, Samsung's Poseidon reference system~\cite{inspur} can support up to 24 Samsung PM1733 NVMe SSDs with a total capacity of up to 368TiB. This capacity is sufficient to support thousands of cache clients for cloud block storage.

\begin{figure}[t]
	\centering
	\includegraphics[width=0.9\columnwidth]{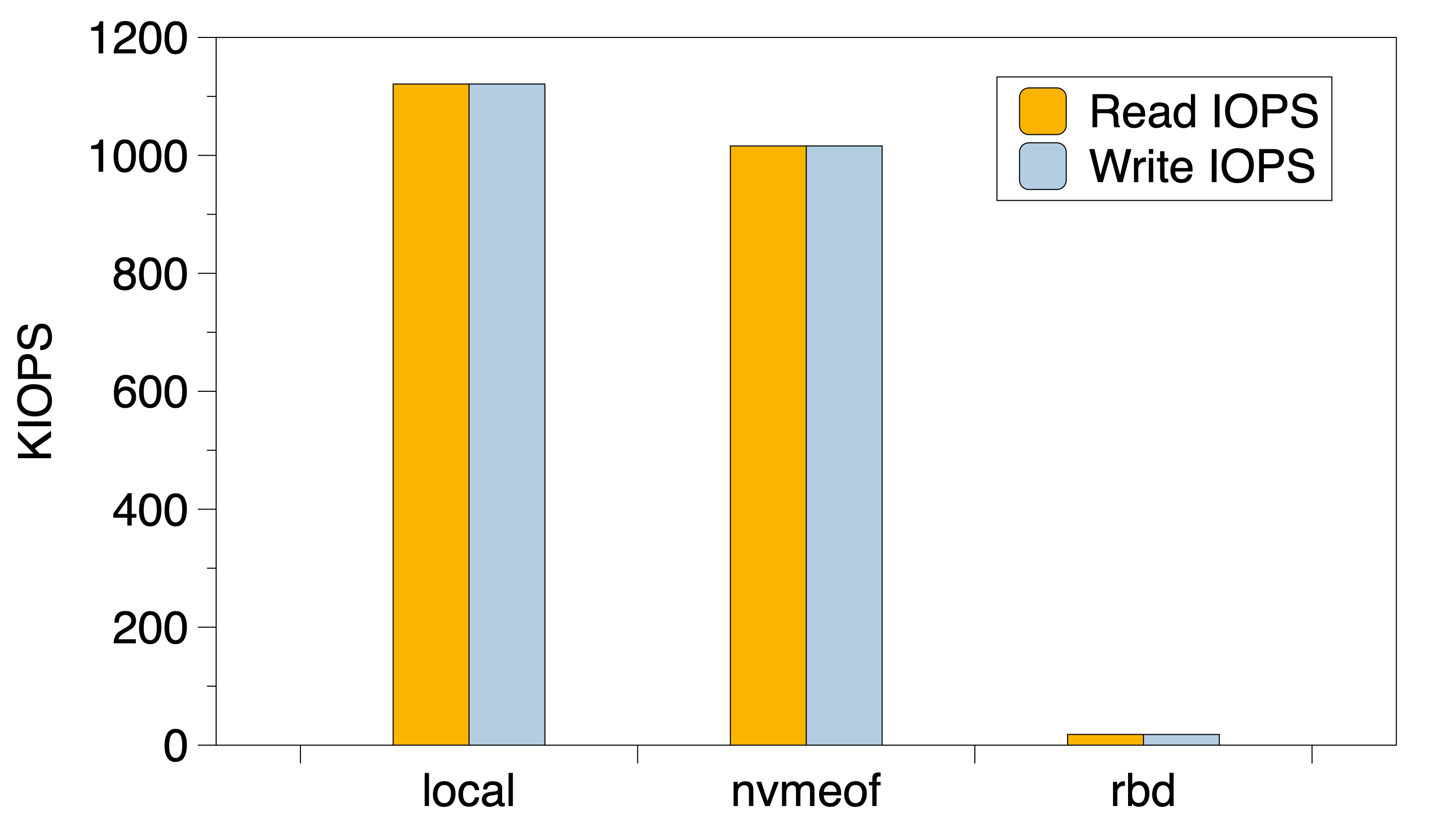}
	\caption{IOPS Comparison of Local SSDs, NVMeoF SSDs, and All-Flash Ceph RBD.}
	\label{fig:compare}
 \vspace{-10pt}
\end{figure}

Figure~\ref{fig:compare} compares the I/O performance of different storage setup: local NVMe SSDs (local), remote NVMeoF SSDs (nvmeof), and remote all-flash Ceph Rados Block Devices (rbd)~\cite{ceph}. Local and nvmeof each consists of four Samsung PM9A3 NVMe SSDs that form a RAID0. Rbd consists of 12 Samsung PM9A3 NVMe SSDs from a 3-node Ceph cluster that form a RAID0. We use local to demonstrate the performance of the local cache, and nvmeof to demonstrate the performance of the disaggregated cache. Rbd is an open-sourced cloud block storage system used to demonstrate the performance of cloud block storage without NVMe SSD caching. We ran the FIO~\cite{fio} benchmark issuing 30 minutes of asynchronous random 4K reads and writes with the same I/O queue depth to different storage setups. We observe that local NVMe SSDs outperform cloud block storage by 60X. Remote SSDs using NVMeoF have comparable performance to local NVMe SSDs with merely a 9\% drop in IOPS.

\subsection{Rack-Scale Cache Management}
A cache block is the minimum unit of cache that can be read from or written to. The block size determines the size of an I/O operation that can be performed. Common cache block sizes range from 512B to 64KiB~\cite{shards, osca, cachededup, cloudcache, qcache}. The choice of cache block size can impact the performance, endurance, and cost of a storage solution by affecting cache hit ratio, I/O volume, and in-memory metadata overhead. Therefore, it's important to select a cache block size that fits the workload best. Smaller cache blocks often have better I/O performance due to the smaller I/O volume, which comes from the smaller cache block allocation and smaller cache miss penalty. However, they may have a lower cache hit ratio because they cannot fully leverage the spatial locality within the application requests~\cite{hennessy2011computer}. 

For a rack-scale cache with hundreds of terabytes of cache space, the large memory footprint for the metadata is another concern for small block sizes. For example, assuming each cache block only requires 40 bytes of memory metadata to provide a source address to cache address mapping (including source address, cache address, a pointer for indexing, and two pointers for LRU)~\cite{cloudcache, memcached}, a 368 TiB cache with 16 KiB cache block size would require 920 GiB of memory footprint, which is difficult to fit in memory, considering memory density grows 10 times slower than SSD density~\cite{memdensity}. 

Large cache blocks, on the other hand, can potentially improve hit ratio~\cite{hennessy2011computer} due to better exploitation of spatial locality. Additionally, the memory footprint reduces linearly with the increased size of the cache block. Take the last example: a 368 TiB cache with 512 KiB cache block size would require merely 29 GiB of memory footprint. However, large cache blocks lead to large cache block allocation and large cache miss penalty which can significantly harm I/O performance. These reasons stop large cache blocks from being applied in reality. Section~\ref{sec:eval} presents a thorough comparison of I/O performance using cache of different cache block sizes. 

\begin{figure}[t]
	\centering
	\includegraphics[width=0.9\columnwidth]{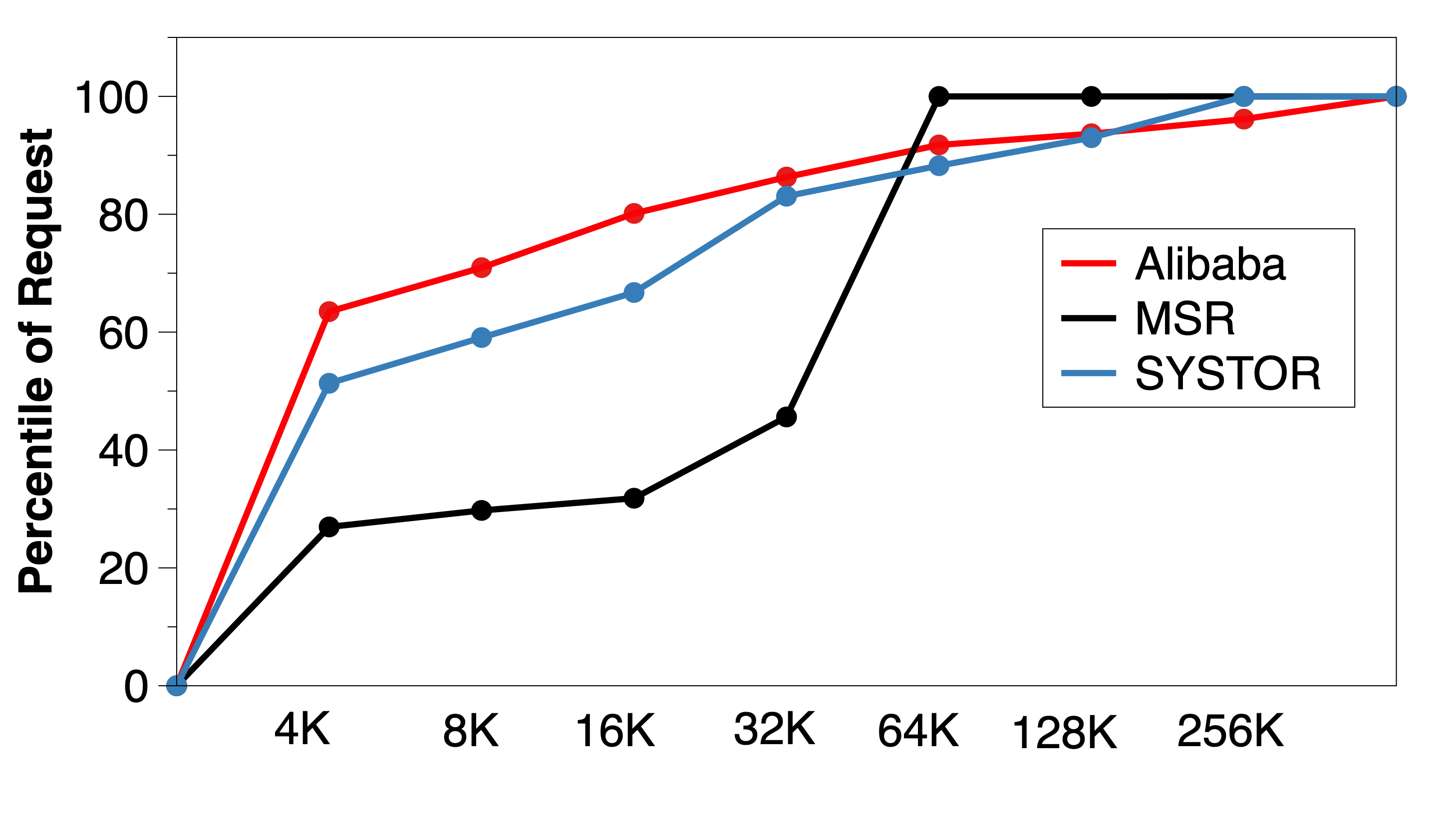}
	\caption{Request Size CDF of different traces }
	\label{fig:cdf}
 \vspace{-5pt}
\end{figure}

The cloud environment is dynamic and changes rapidly over time with varying workloads. Some workloads involve small requests, such as those from transactional databases, while others have large requests, such as those from multimedia systems. We conducted an analysis of request size cumulative distribution functions (CDF) from three real-world traces: Alibaba block I/O Traces (\textit{alibaba})~\cite{li2020depth}, MSR Cambridge Traces (\textit{msr})~\cite{narayanan2008write}, and Systor '17 Traces (\textit{systor})~\cite{lee2017understanding} (detailed information about the traces is presented in Section~\ref{sec:eval}). Figure~\ref{fig:cdf} shows the results. We observe that the distribution of request sizes varies across the traces. For \textit{alibaba} and \textit{systor}, more than half of the requests are smaller than or equal to 4KiB. For \textit{msr}, more than half of the requests are larger than 32KiB. 
Based on the above observations, a traditional fix-sized block cache is insufficient for today's complex cloud environment. Instead, we design an adaptive cache that can adapt the cache block size to different cloud workloads which is elaborated in Section~\ref{sec:design}. 

\begin{figure}[t]
	\centering
	\includegraphics[width=1.0\columnwidth]{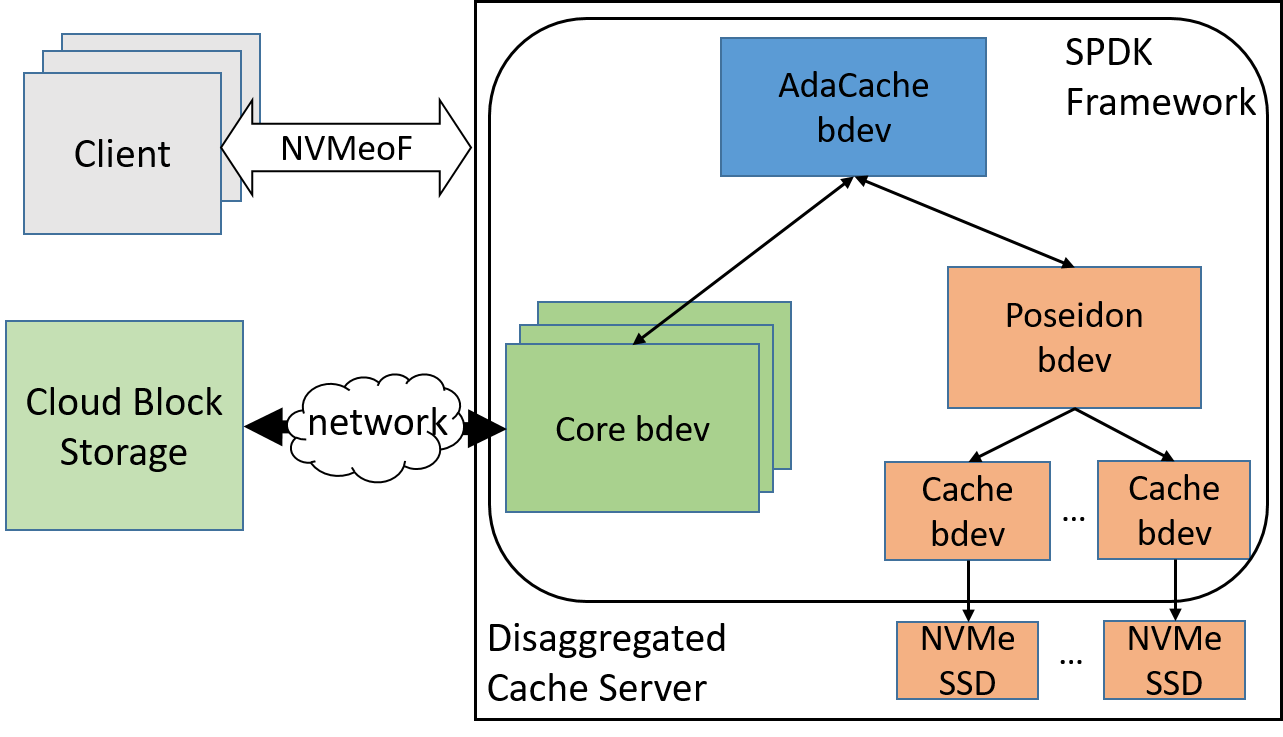}
	\caption{Disaggregated Cache Architecture}
	\label{fig:testbed}
 \vspace{-10pt}
\end{figure}

\begin{figure*}[t]
	\centering
	\includegraphics[width=2\columnwidth]{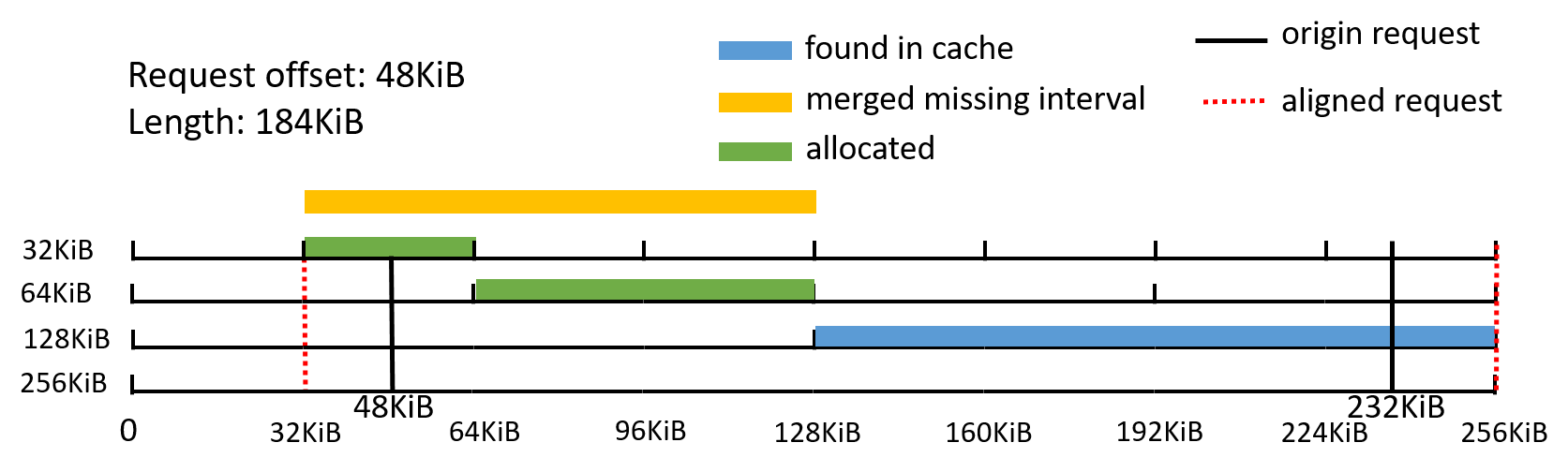}
	\caption{Adaptive Cache Block Allocation}
	\label{fig:alloc}
 \vspace{-10pt}
\end{figure*}

\subsection{Implementation}
\label{sec:imp}
AdaCache extends PoseidonOS~\cite{poeidon}, a userspace software-defined storage (SDS) solution providing high-throughput and low-latency flash storage virtualization with capacity elasticity and data protection (RAID), to offer rack-scale disaggregated cache service for cloud block storage. It is implemented as a virtual block device (bdev) module~\cite{bdev} using the SPDK framework. By using a virtual bdev module, AdaCache can be seamlessly integrated with a wide range of cloud block storage bdevs, enabling compatibility with existing storage systems. 

Figure~\ref{fig:testbed} illustrates the architecture of AdaCache. Each local NVMe SSD is represented by a cache bdev in the SPDK framework. All the cache bdevs are managed by PoseidonOS to offer a large virtualized disaggregated cache space to AdaCache. Each virtual drive in the cloud block storage is represented by a core bdev.
AdaCache claims the cache and core bdevs and redirects I/Os between them with no requirement for knowledge of the I/O and network protocol specifics of the underlying bdevs.
AdaCache uses GLib's~\cite{glib} hash table implementation for the in-memory key-value stores.

\Section{Adaptive Cache Block Size}
\label{sec:design}
\algnewcommand{\Inputs}[1]{%
  \State \textbf{Inputs:}
  \Statex \hspace*{\algorithmicindent}\parbox[t]{.8\linewidth}{\raggedright #1}
}
\algnewcommand{\Outputs}[1]{%
  \State \textbf{Output:}
  \Statex \hspace*{\algorithmicindent}\parbox[t]{.8\linewidth}{\raggedright #1}
}
\algnewcommand{\Remarks}[1]{%
  \State \textbf{Remarks:}
  \Statex \hspace*{\algorithmicindent}\parbox[t]{.8\linewidth}{\raggedright #1}
}
\algnewcommand\algorithmicforeach{\textbf{for each}}
\algdef{S}[FOR]{ForEach}[1]{\algorithmicforeach\ #1\ \algorithmicdo}
\vspace{-5pt}
\subsection{Fix-sized Cache Allocation}
\label{sec:fix-alloc}
Traditional fix-sized cache block allocation has three major steps: address alignment, address lookup, and cache block allocation. Address alignment aligns the offset of the original I/O requests to the aligned offset based on cache block size. Assume $R^{}_o$ is the request offset, $B$ is the cache block size, and $A^{}_o$ is the aligned offset. $A^{}_o$ is computed using the following Equation~\ref{eq1}.
\begin{equation}
\begin{aligned} 
\label{eq1}
A^{}_o = floor(R^{}_o / B) * B \\
\end{aligned} 
\end{equation}
For example, a read request with offset 33KiB using 32KiB as cache block size aligns to aligned offset 32KiB.

During address lookup, the aligned offset is used as the key to look up the cache address in an in-memory key-value store. In case of a read cache hit, data is read from the cache address directly. Otherwise, a new cache block is allocated and data is read from the backend storage and cached to the newly allocated cache block. 

In case of a write cache miss, data is first read from the backend storage and cached to a newly allocated cache block. If the cache uses write-back policy, data is written to the cache block and dirty cache blocks are written back to the backend storage periodically or when they are replaced from the cache. If the cache uses write-through policy, data is written to the cache block and backend storage simultaneously to maintain data consistency. When the cache becomes full, a replacement algorithm such as Least Recently Used (LRU) or Least Frequently Used (LFU) is used to determine which data to replace before allocation happens.

\subsection{Variable-Sized Cache Allocation}
\label{sec:var-alloc}

Cloud workloads are dynamic in nature, and therefore, the cache system should be able to adapt itself to different workloads that may have varying request sizes. For small requests, small cache blocks are deemed sufficient while large cache blocks may cache unnecessary data, resulting in cache pollution and increased I/O volumes. Conversely, for large requests, large cache blocks can reduce the number of I/Os between the cache and the cloud block storage, and can also reduce the metadata memory overhead. AdaCache uses adaptive cache block allocation which allocates different sizes of cache blocks based on the request size.

AdaCache first generates a list of missing intervals for all the parts of the request that are missing in the cache. 
As shown in Figure~\ref{fig:cdf}, a request can be larger than 256KiB and cover multiple cache blocks.
AdaCache determines the aligned range of the request by aligning the request offset and end address (offset + length) to the smallest block size and iterates through the request to find out all the missing intervals. 

Because the cache employs variable cache block sizes, it needs to check the in-memory key-value store of every block size to find out if any part of the request is cached under each block size. Figure~\ref{fig:alloc} illustrates an example where a request at offset 48KiB with length 184KiB on a cache that employs cache block sizes of 32KiB, 64KiB, 128KiB, and 256KiB. In this example, the latter part of the request (from 128KiB to 232KiB) is cached under the 128KiB block size. The aligned request range is from 32KiB to 256KiB.

Within the request range, AdaCache starts the search from the smallest cache block size (32KiB in the example), and checks if the current address is cached under any of the cache block sizes. 
AdaCache first aligns the current address to different cache block sizes using Equation~\ref{eq1}. For the example, the aligned offsets are 32KiB, 0, 0, and 0 for the cache block sizes of 32KiB, 64KiB, 128KiB, and 256KiB respectively. It then uses these aligned offsets to search the in-memory key-value store of each cache block size. If the result is all misses, then it knows that the current address with the smallest cache block size (the interval between 32KiB and 64KiB in the example) is not cached, and it adds the interval to the list of missing intervals. AdaCache merges missing intervals if they are contiguous to allocate the largest possible cache block for the intervals.
AdaCache then moves on to the next address covered by the request (64KiB in the example) and repeats the above process. 
After checking the whole request, AdaCache gets a complete list of missing intervals. In the example, the interval from 32KiB to 128KiB is missing in the cache. Algorithm~\ref{ag1} presents the pseudo-code of the missing intervals generation.

\begin{algorithm}
  \caption{Missing Intervals Generation}
  \label{ag1}
  \begin{algorithmic}[1]
    \Remarks {\strut$B_n,\ldots$ $B_1$: block size from large to small  \\
        $H_{B}$: hash table for block size $B$ \\ 
        $A_{B}(O)$: align offset $O$ using block size $B$\\
        $M_{AP}(B, E)$: merge offset interval $\{B, E\}$ to $MissingIntervals$}
    \Inputs {$O$: request offset in bytes \\
            $L$: request length in bytes}
    \Outputs {$MissingIntervals$: a list of missing cache blocks}
    \State $MissingIntervals \gets \{\}$
    \State $begin \gets A_{B_1}(O)$     
    \State $end \gets A_{B_1}(O + L) + B_{1}$
    \While {$begin \neq end$}
        \State $hit \gets \textbf{false}$
        \For{$B \gets B_1,\ldots B_n$}
            \State $begin\_aligned = A_{B}(begin)$  
            \If {$begin\_aligned \in H_B$}  
                \State $begin \gets begin\_aligned + B$
                \State $hit \gets \textbf{true}$
                \State \textbf{break}
            \EndIf
        \EndFor
        \If {$hit \neq \textbf{true}$}
            \State $M_{AP}(begin, begin + B_1)$ 
            \State $begin \gets begin + B_1$
        \EndIf
    \EndWhile
    \State \textbf{return $MissingIntervals$}
  \end{algorithmic}
\end{algorithm}

\begin{algorithm}
  \caption{Greedy Cache Block Allocation}
  \label{ag2}
  \begin{algorithmic}[1]
    \Remarks {\strut$B_1,\ldots$ $B_n$: block size from small to large  \\
        $H_{B}$: hash table for block size $B$ \\ 
        $A_{B}(O)$: align offset $O$ on block size $B$ \\
        $BA(I)$: the begin address of interval $I$ \\
        $EA(I)$: the end address of interval $I$}
    \Inputs {$MissingIntervals$: a list of cache blocks to allocate}
    \vspace{+5pt}
    \ForEach{$I \in MissingIntervals$}
        \State $begin\gets BA(I)$
        \State $end \gets EA(I)$
        \While{$begin \neq end$}
            \For{$B \gets B_n,\ldots B_1$}  
                \If {$begin \neq A_{B}(begin)$}
                    \State \textbf{continue}
                \EndIf
                \If {$B > end - begin$}
                    \State \textbf{continue}
                \EndIf
                \State $H_B \gets begin \cup H_B$   \Comment{allocate cache block}
                \State $begin \gets begin + B$
            \EndFor
        \EndWhile
    \EndFor
  \end{algorithmic}
\end{algorithm}

For each missing interval in the list, AdaCache tries to allocate using the largest possible cache block size. This greedy allocation ensures that AdaCache reduces the number of allocated cache blocks and I/O counts. 
To determine if a block size is suitable for the missing interval, AdaCache makes sure the cache block is within the range of the missing intervals because the addresses that go beyond these intervals may have been cached.

In the example, AdaCache first checks how to allocate for the interval from 32KiB to 128KiB. The largest possible cache block for this interval is actually 32KiB, because all the larger cache blocks start beyond this interval.
For the remaining missing interval from 64KiB to 128KiB, the largest possible cache block is 64KiB, because the interval from 64KiB to 128KiB is within the range of the missing interval (64KiB to 128KiB).
Therefore, at the end of this greedy allocation process, AdaCache caches two blocks that include one 32KiB cache block from 32KiB to 64KiB and one 64KiB cache block from 64KiB to 128KiB. Algorithm~\ref{ag2} presents the pseudo-code of the greedy cache block allocation.

Assuming $N$ is the request length, $M$ is the number of different cache block sizes, and $K$ is the total number of cache blocks in the cache, the algorithm's time complexity of fix-sized and adaptive cache block allocation have upper bounds of O($K*N$) and O($K*N*M$), respectively. In practice, $M$ is set to a constant value, such as 4 in Figure~\ref{fig:alloc} where the time complexity can be approximated as O($K*N$), which is equivalent to the fix-sized cache block allocation. 
The space complexity of the algorithm is identical to the fix-sized cache block allocation, which is O($K$).

\begin{figure}[t]
	\centering
	\includegraphics[width=1.0\columnwidth]{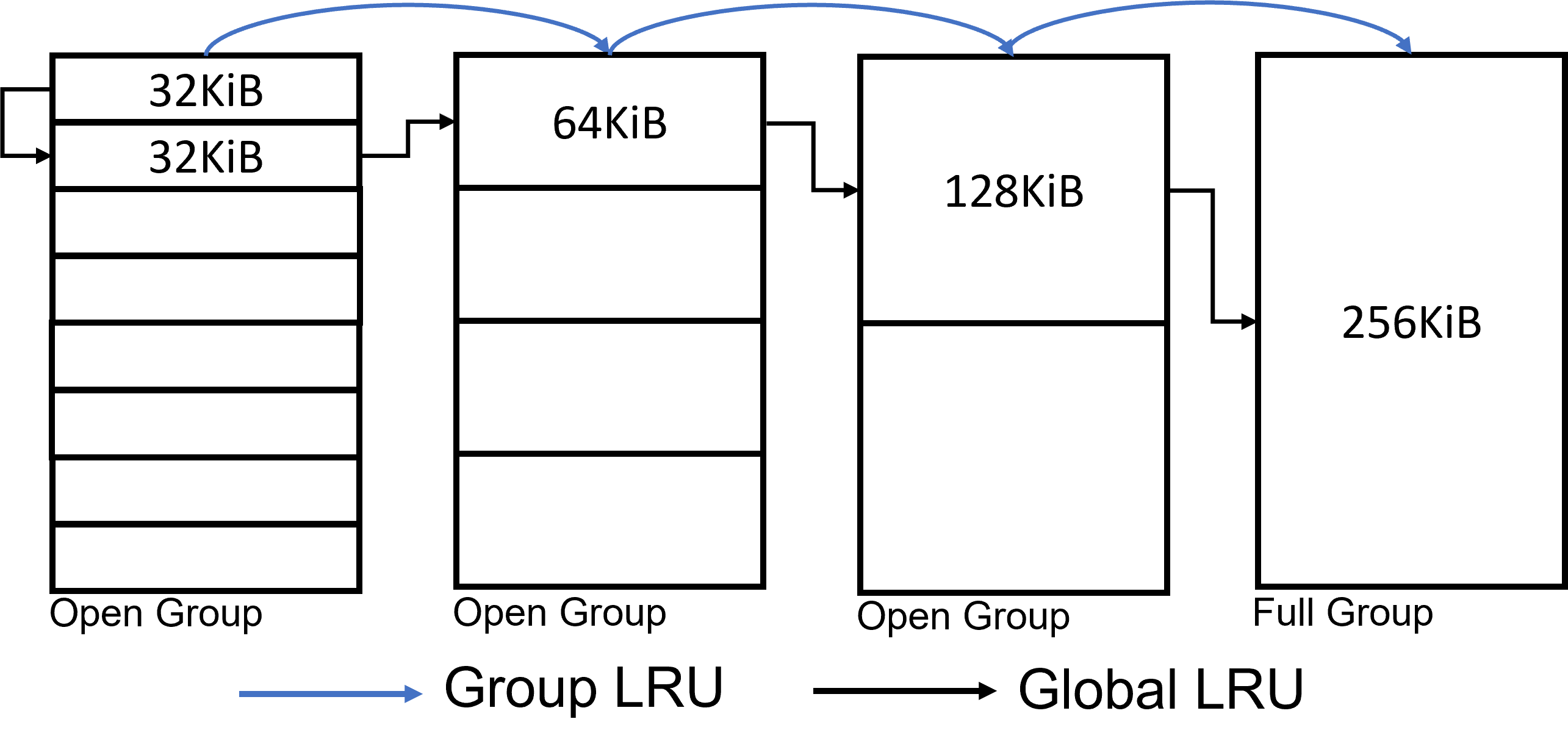}
	\caption{Group-Based Cache Organization}
	\label{fig:group}
 \vspace{-10pt}
\end{figure}

\vspace{-5pt}
\subsection{Group-Based Cache Organization}
\vspace{-3pt}
Adaptive cache block allocation is an effective technique that can leverage both small and large blocks, making it suitable for dynamic cloud workloads. However, it incurs fragmentation. When the cache becomes full and adaptive cache blocks get allocated, the cache space is divided into non-contiguous variable-sized pieces. When large requests come, the replacement of smaller blocks can generate many scattered small holes and it is hard to fit a large cache block in. 

To address the issue of fragmentation, AdaCache utilizes the concept of slab allocator~\cite{slab, memcached}, which involves grouping cache blocks of the same size together into identical-sized groups. Cache blocks belonging to the same group are stored physically adjacent to each other in the cache. Consequently, when the cache is full, a whole group is replaced, creating a contiguous piece of cache space for cache block allocation. 

AdaCache chooses the largest cache block size as the group size. In this way, replacement of a whole group can free just enough cache space for the largest cache block allocation. In the case of small block allocation, the replacement of a whole group creates an open group that can be used to allocate many cache blocks of that block size. Figure~\ref{fig:group} illustrates an example of the group-based cache organization. The cache block sizes are 32KiB, 64KiB, 128KiB, and 256KiB and the group size is 256KiB. There are three open groups storing 32KiB, 64KiB, and 128KiB cache blocks, respectively, and one full group storing a 256KiB cache block.  

When allocating a cache block, AdaCache checks if the cache is full. If it is not, the allocator examines if there is an open group with the same block size. If such a group exists, the block is allocated from the open group. If there is no such open group, AdaCache creates a new one and allocates the cache block from there. If the cache is full, AdaCache replaces an entire group and follows the above procedure. Assume $M$ is the number of different cache block sizes, there are a maximum of $M$ open groups kept in the cache at any given time, and it does not waste significant cache space. For example, in Figure~\ref{fig:group}, at most $4$ 256KiB open groups are kept in the cache and used to allocate cache blocks for coming requests.
 
\subsection{Two-Level Cache Replacement}
Following group-based cache organization, AdaCache uses a group-based LRU replacement policy that links all the groups together for cache replacement. When a cache block is accessed, the group that contains the cache block is promoted to the head of the group-based LRU list. When the cache is full, AdaCache replaces the group that is at the tail of the LRU list. Although each cache miss may trigger a write-back I/O of the whole group to be evicted, the I/O volume is smaller than that of using large fix-sized cache blocks. Every time a whole group is evicted, all of its space is freed up at once in the cache and can be used to store a number of small cache blocks from future requests. 

One potential drawback of the group-based replacement policy is that it may retain cold blocks that are in the same group as the frequently accessed hot blocks in the cache, leading to cache pollution. To alleviate the problem, AdaCache incorporates a global cache block LRU replacement policy in addition to the group-based replacement policy. Figure~\ref{fig:group} illustrates the two-level LRU lists.

All the cache blocks are linked using a global LRU list. When AdaCache tries to allocate a new cache block in case of a full cache, it first checks the tail of the global LRU list. If the tail cache block has the same size as the new cache block, AdaCache replaces it and promotes both the cache block and its group to the head of the LRU lists. If the size mismatches, AdaCache uses group-based LRU replacement policy to replace a whole group.  
The use of two-level cache replacement does not incur high lock contention overhead when the cache is accessed in parallel as AdaCache leverages the lockless design of modern high performance storage framework~\cite{spdk}.

\Section{Evaluation}
\label{sec:eval}

We evaluate the performance of AdaCache using both the simulation and prototype following the design and implementation described in Section~\ref{sec:design}. 

\begin{table}[t]
	\centering
	\caption{Specifications Of The Testbed.}
	\small
	\tabcolsep=0.11cm
	\begin{tabular}{|c|c|c|c|c|c|}
		\hline
		\textbf{Server}&\textbf{CPU}&\textbf{DRAM}	&\textbf{SSD} &\textbf{OS} &\textbf{Software}\\    
		\hline
		   & 2x Intel & &  &  &   \\
		                Client& Platinum & 384GB &   & Ubuntu & Replayer / \\
    		            &8260  &     DDR4    &  / &18.04 &  Simulator \\
                  &96 cores & &  & &  \\
		\hline
   & 2x Intel & &4x Samsung & &   \\
		               Disaggregated  & Platinum & 384GB & PM9A3  & Ubuntu & Poseidon  \\
    		          Cache Server &8260 &     DDR4    &   PCIe Gen3   & 20.04 & OS v0.11      \\
                  &96 cores & & 3.84TB& &  \\
                  \hline
                    &  & &4x Samsung &  &  \\
		                3-node& 2x AMD & 512GB & PM9A3  & Ubuntu & Ceph \\
    		           Ceph RBD &EPYC    &     DDR4    &   PCIe Gen4  & 20.04 &  Quincy     \\
                  &7702 & & 3.84TB & &  \\
\hline	
 \end{tabular}
	\label{table:table1}
 \vspace{-10pt}
\end{table}

\begin{figure}[t]
     \begin{subfigure}[b]{0.5\textwidth}
	\includegraphics[width=0.9\textwidth]{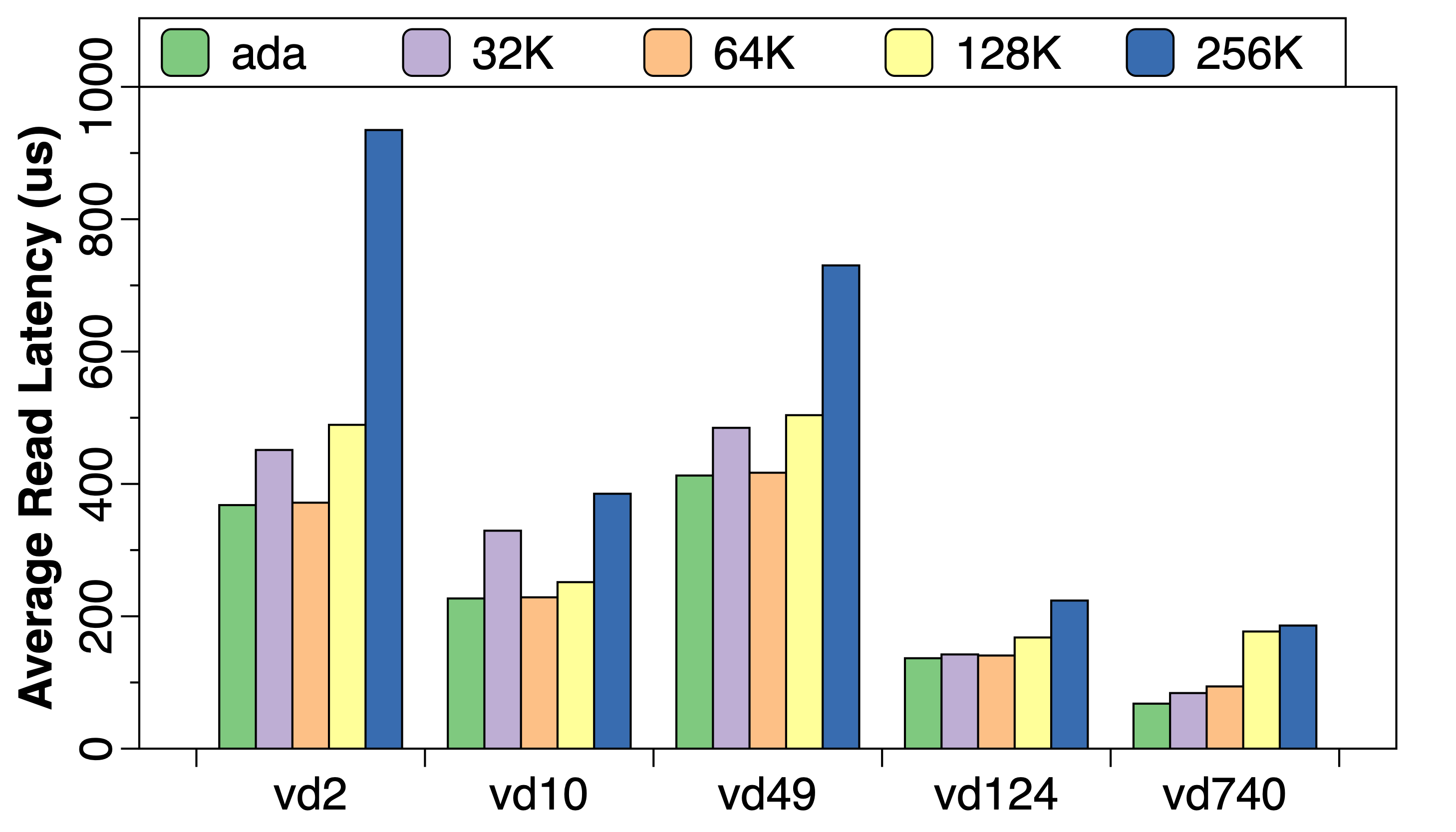}
	 \caption{Read Latency}
	 \label{fig:alibabarl}
\end{subfigure}
   \hfill
 \begin{subfigure}[b]{0.5\textwidth}
    \includegraphics[width=0.9\textwidth]{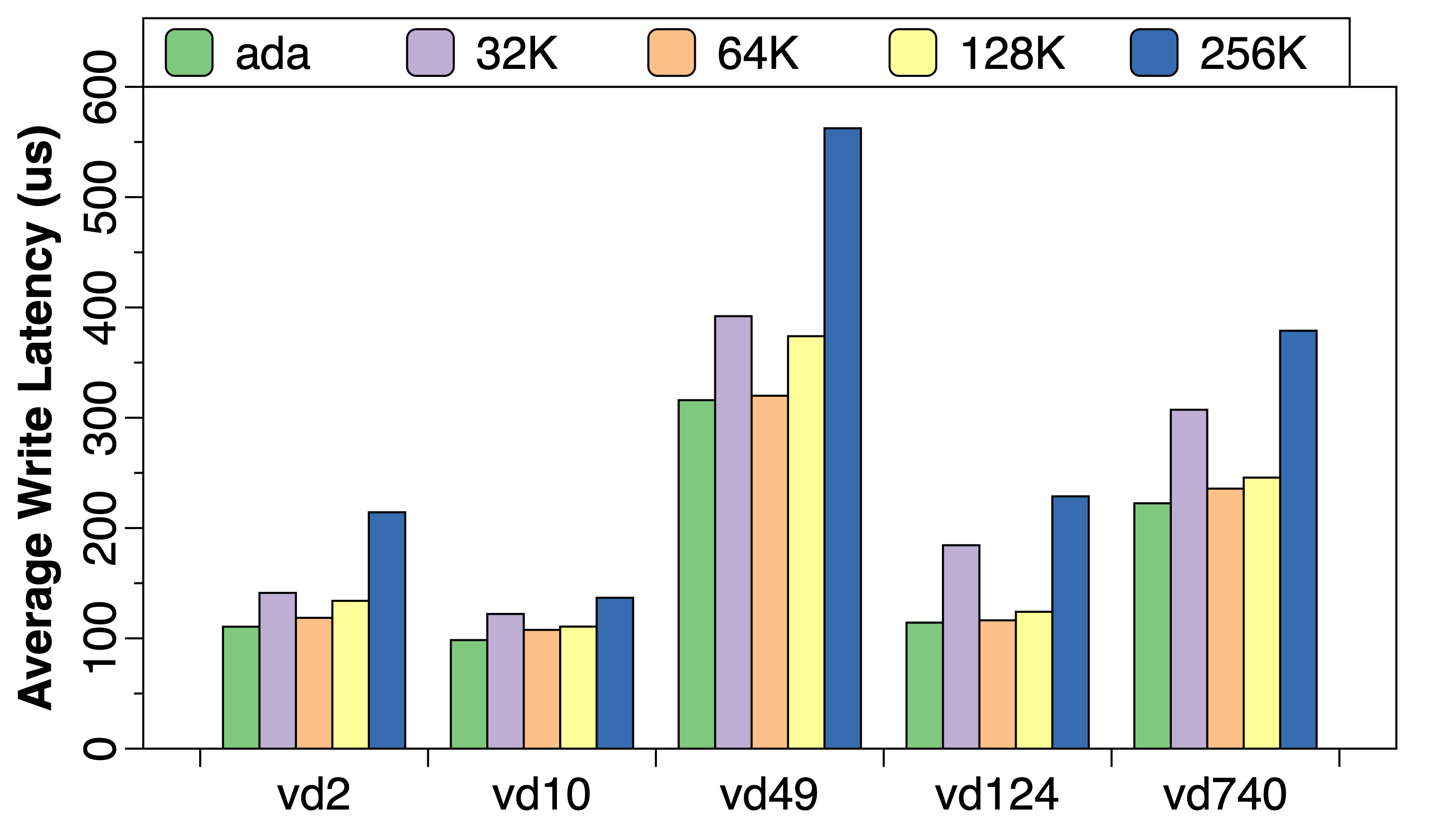}
    \caption{Write Latency}
    \label{fig:alibabawl}
 \end{subfigure}
\caption{I/O Latency for Alibaba Trace Replay}
\label{fig:fig1}
\vspace{-10pt}
\end{figure}

\begin{figure}[t]
     \begin{subfigure}[b]{0.5\textwidth}
	\includegraphics[width=0.9\textwidth]{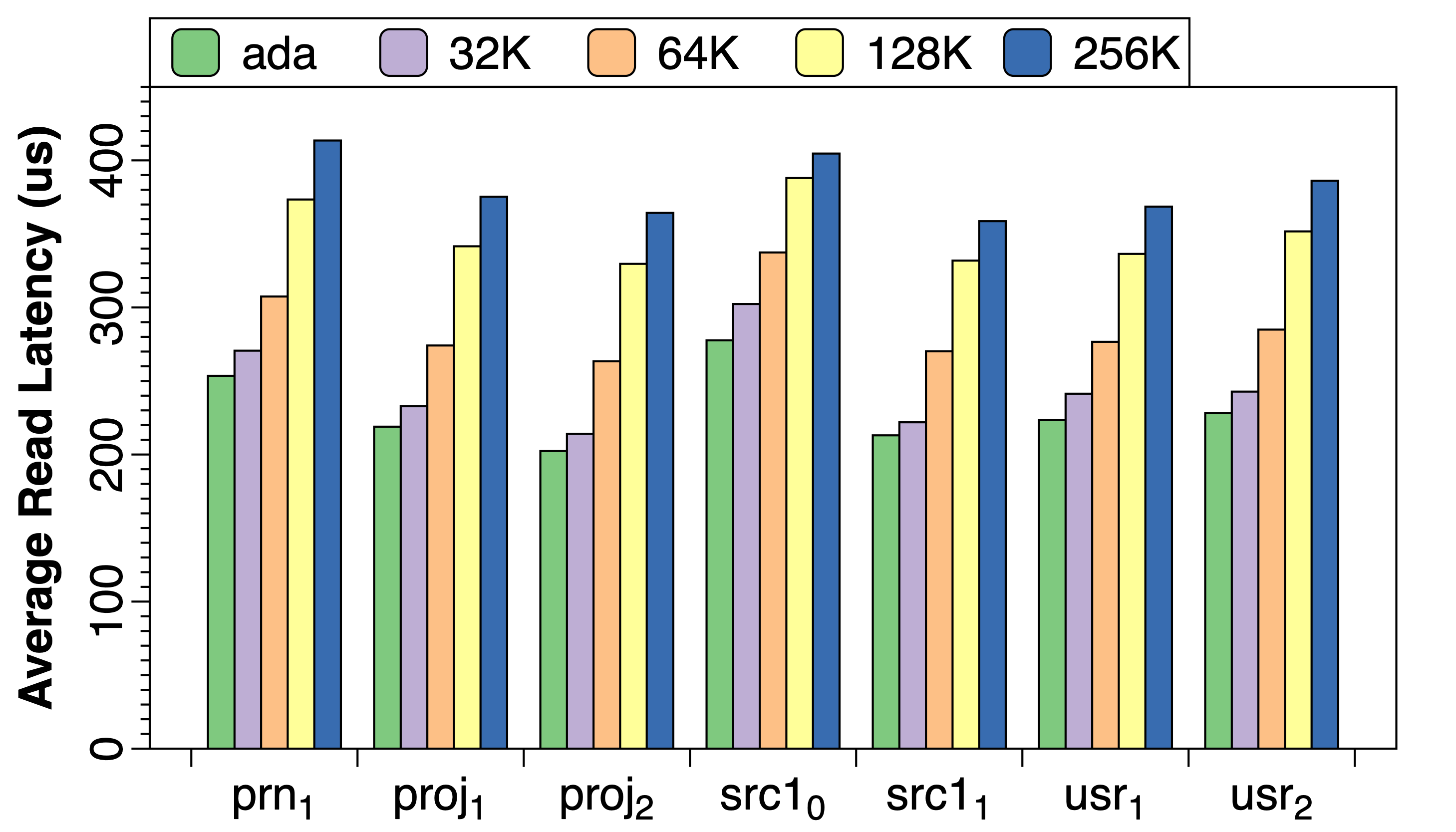}
	 \caption{Read Latency}
	 \label{fig:msrrl}
\end{subfigure}
   \hfill
 \begin{subfigure}[b]{0.5\textwidth}
    \includegraphics[width=0.9\textwidth]{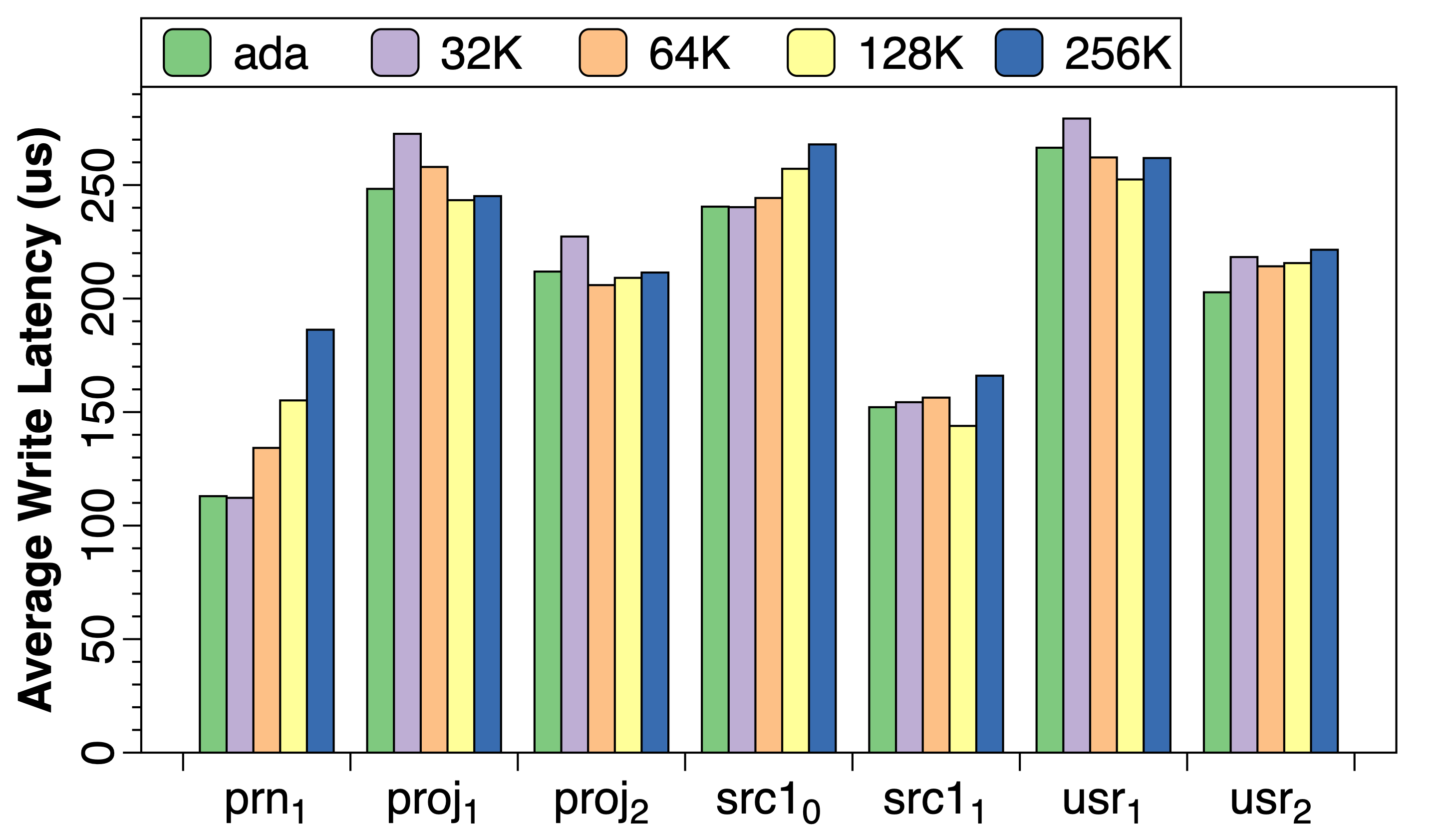}
    \caption{Write Latency}
    \label{fig:msrwl}
 \end{subfigure}
\caption{I/O Latency for Msr Trace Replay}
\label{fig:fig2}
\vspace{-10pt}
\end{figure}



\begin{table}[t]
	\centering
	\caption{Trace Segments Statistics.}
	\small
	\tabcolsep=0.41cm
	\begin{tabular}{|c|c|c|c|}
		\hline
		&\textbf{\textit{alibaba}}&\textbf{\textit{msr}}&\textbf{\textit{systor}}\\    
		\hline
  \#Reads           & 24.5M   & 61M    & 40.7M  \\
  \hline
\#Writes          & 25.5M   & 9M     & 19.3M \\
\hline
Read Traffic GiB  & 607.3   & 2416.8 & 1109.2  \\
\hline
Write Traffic GiB & 375.9   & 207.2  & 271.9  \\
\hline	
 \end{tabular}
 \hspace{-10pt}
\label{table:table2}
\vspace{-10pt}
\end{table}

\textbf{Testbed Setup.}
The testbed consists of three components which are the client, the disaggregated cache server, and the cloud block storage cluster. The client issues the I/O workloads to the disaggregated cache server through NVMeoF RDMA using a 100Gbps NIC. The disaggregated cache server runs AdaCache and provides the cloud block storage with NVMe SSD caching through the network using another 100Gbps NIC. The disaggregated cache server is configured as RAID0 using PoseidonOS consisting of four NVMe SSDs. The cloud block storage is a three-node Ceph cluster with Ceph Rados Block Devices (RBDs). The specs for each component are shown in Table~\ref{table:table1}.

\textbf{Workloads.}
We considered the following three real-world block I/O traces to provide a comprehensive evaluation:
\begin{itemize}
\item Alibaba block I/O Traces~\cite{li2020depth} (\textit{alibaba}): \textit{alibaba} is collected from an elastic block service cluster of Alibaba Cloud and it contains I/Os from 1000 virtual disks. Among them, we picked 5 virtual disks (vd2, vd10, vd49, vd124, and vd740) that have a large amount of I/O volumes for trace replay. We replayed the first 10 million I/O requests issued to the 5 virtual disks concurrently. Requests to vd2 and vd740 are write-dominant while I/Os to vd10 and vd124 are read-dominant. Vd49 has a similar amount of read and write I/Os. 

\item MSR Cambridge Traces~\cite{narayanan2008write} (\textit{msr}): \textit{msr} is block-level traces collected from Microsoft Research enterprise data centers and it contains I/Os from 13 servers. Among them, we picked traces from seven drives (prn\_1, proj\_1, proj\_2, src1\_0, src1\_1, usr\_1, and usr\_2) that have more than 10 million I/Os. We replayed the first 10 million I/Os issued to the 7 servers concurrently. All the \textit{msr} traces are read-dominant. 

\item Systor '17 Traces~\cite{lee2017understanding} (\textit{systor}): \textit{systor} is collected from an enterprise Virtual Desktop Infrastructure (VDI) which contains I/Os from 300 VMs. All these VMs share 6 storage logical unit numbers (LUN). We replayed the first 10 million I/Os issued to the 6 LUNs concurrently. All the \textit{systor} traces are read-dominant. 
\end{itemize}

For trace segments replay, the cache employs a write-back policy and we can leverage related work~\cite{koller2013write} to ensure cache consistency. Trace segments are replayed using pread() and pwrite() to issue direct I/Os to different target devices in parallel according to the trace. Each target device consists of 1 TiB Ceph RBD as the backend storage and 10\% of each trace's total working set size (WSS) as the cache size. Table~\ref{table:table2} shows the statistics of the trace segments that we use for replay. We also replay the entire traces using a simulator with the same implementation as the AdaCache prototype to show metrics from the whole trace simulation. In the evaluation, the cache block sizes used by AdaCache are 32KiB, 64KiB, 128KiB, and 256KiB. We compare AdaCache to fix-sized disaggregated caches with these four cache block sizes. Each experiment is repeated three times and we show the average results here. Due to the space limit, we only show evaluation results that are representative of all results.

\begin{figure}[t]
     \begin{subfigure}[b]{0.5\textwidth}
	\includegraphics[width=0.9\textwidth]{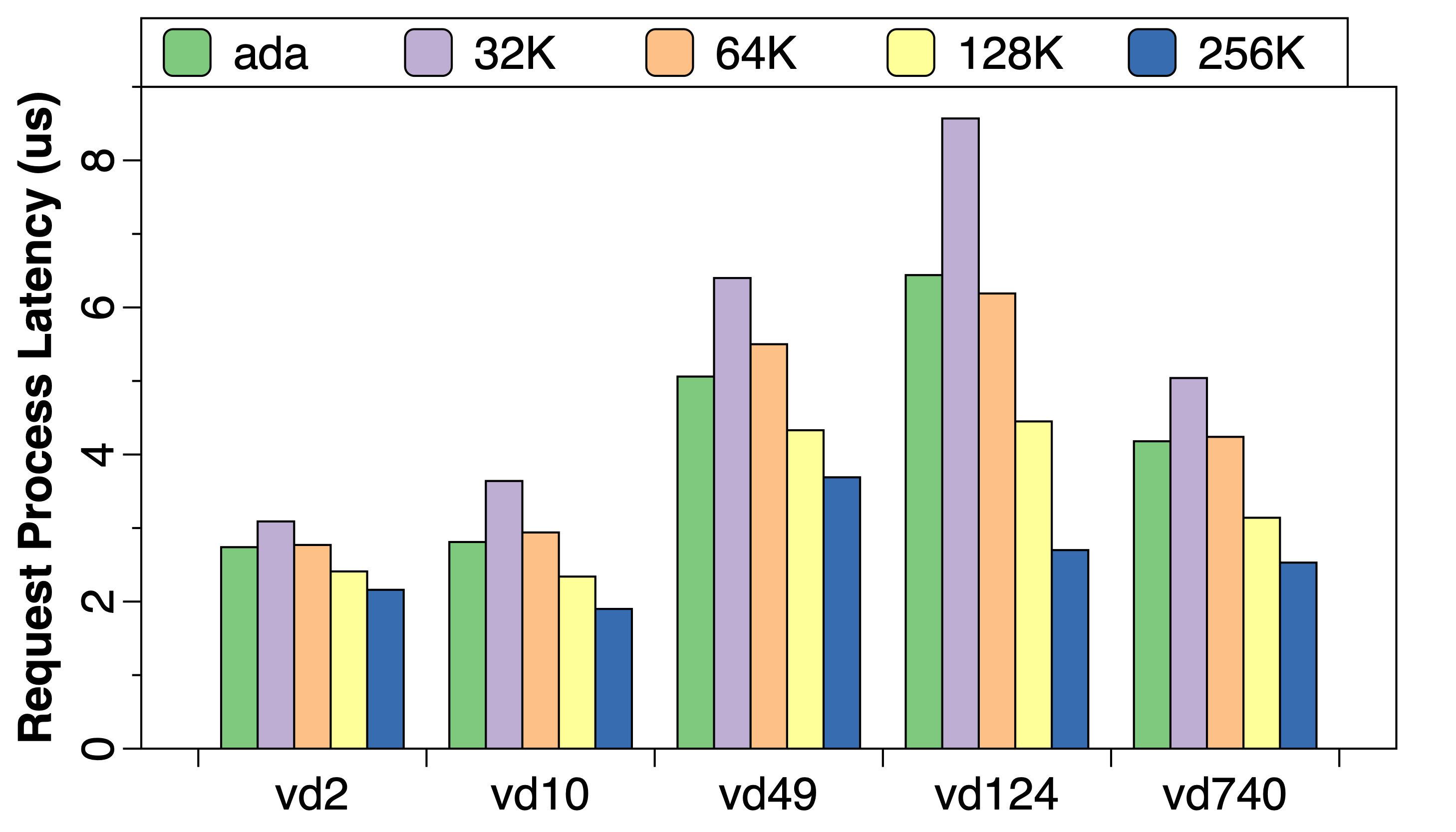}
	 \caption{Alibaba Trace Replay}
	 \label{fig:alibabapl}
\end{subfigure}
   \hfill
 \begin{subfigure}[b]{0.5\textwidth}
    \includegraphics[width=0.9\textwidth]{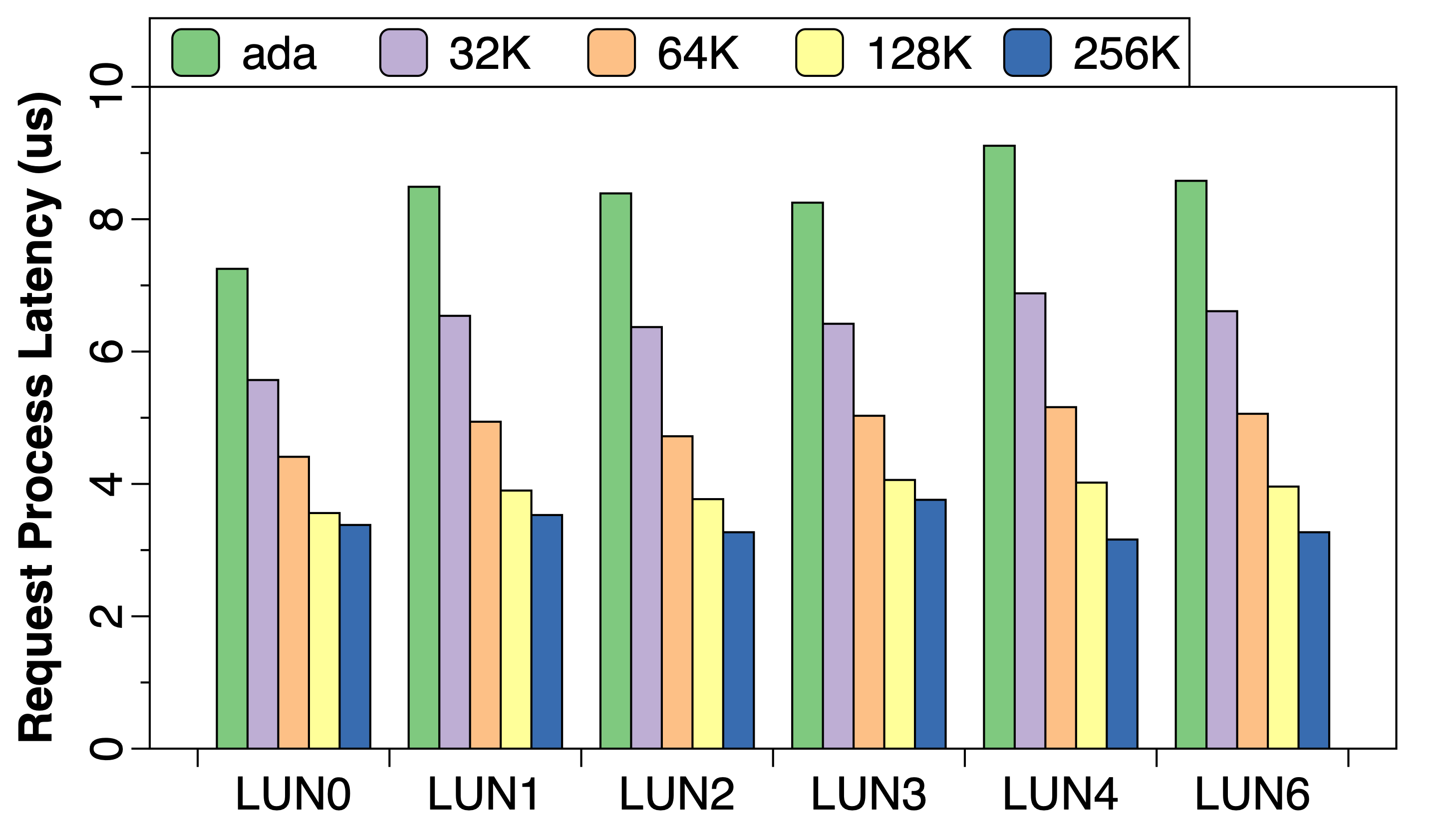}
    \caption{Systor Trace Replay}
    \label{fig:systorpl}
 \end{subfigure}
\caption{Request Processing Latency}
\label{fig:fig3}
\vspace{-10pt}
\end{figure}

\subsection{I/O performance}
\textbf{I/O Latency.}
Figure~\ref{fig:fig1} shows the average read and write latency from \textit{alibaba} trace replay. Results reveal that AdaCache has the best overall read and write latency compared to fix-sized caches with different trace segments. For read latency, AdaCache improves it by 19\% for trace segment vd740 compared to 32KiB cache and 63\% compared to 256KiB cache. For write, AdaCache has an improvement of 9\% for trace segment vd10 compared to 64KiB cache and 50\% for vd124 compared to 256KiB cache. Figure~\ref{fig:fig2} shows the read and write latency from \textit{msr} trace replay. AdaCache also improves the read latency by 7\% compared to 32KiB cache for usr\_1 and 44\% compared to 256KiB cache for proj\_2. For write latency, AdaCache can improve it by 9\% compared to 32KiB cache for proj\_1 and 39\% compared to 256KiB cache for prn\_1. 

Comparing the two traces' latency results from fix-sized caches, \textit{alibaba} mostly has the best read and write performance when using a 64KiB cache. \textit{Msr} has the best read performance when using a 32KiB cache. For write, different cache block sizes perform differently for different trace segments. For example, trace segment prn\_1 has the best write performance using 32KiB cache while trace segment proj\_1 performs the best using 128KiB cache. This also proves that a fix-sized cache cannot provide optimal performance for different cloud workloads. Of the two traces, AdaCache outperforms all the fix-sized caches in both read and write. Although AdaCache has similar I/O volumes as 32KiB cache (discussed later in Section~\ref{sec:io}), it is achieving better performance because of the adaptiveness of AdaCache which allocates large cache blocks for large requests. These large cache blocks have reduced the number of I/Os and can therefore improve the performance.   


\begin{figure}[t]
     \begin{subfigure}[b]{0.5\textwidth}
	\includegraphics[width=0.9\textwidth]{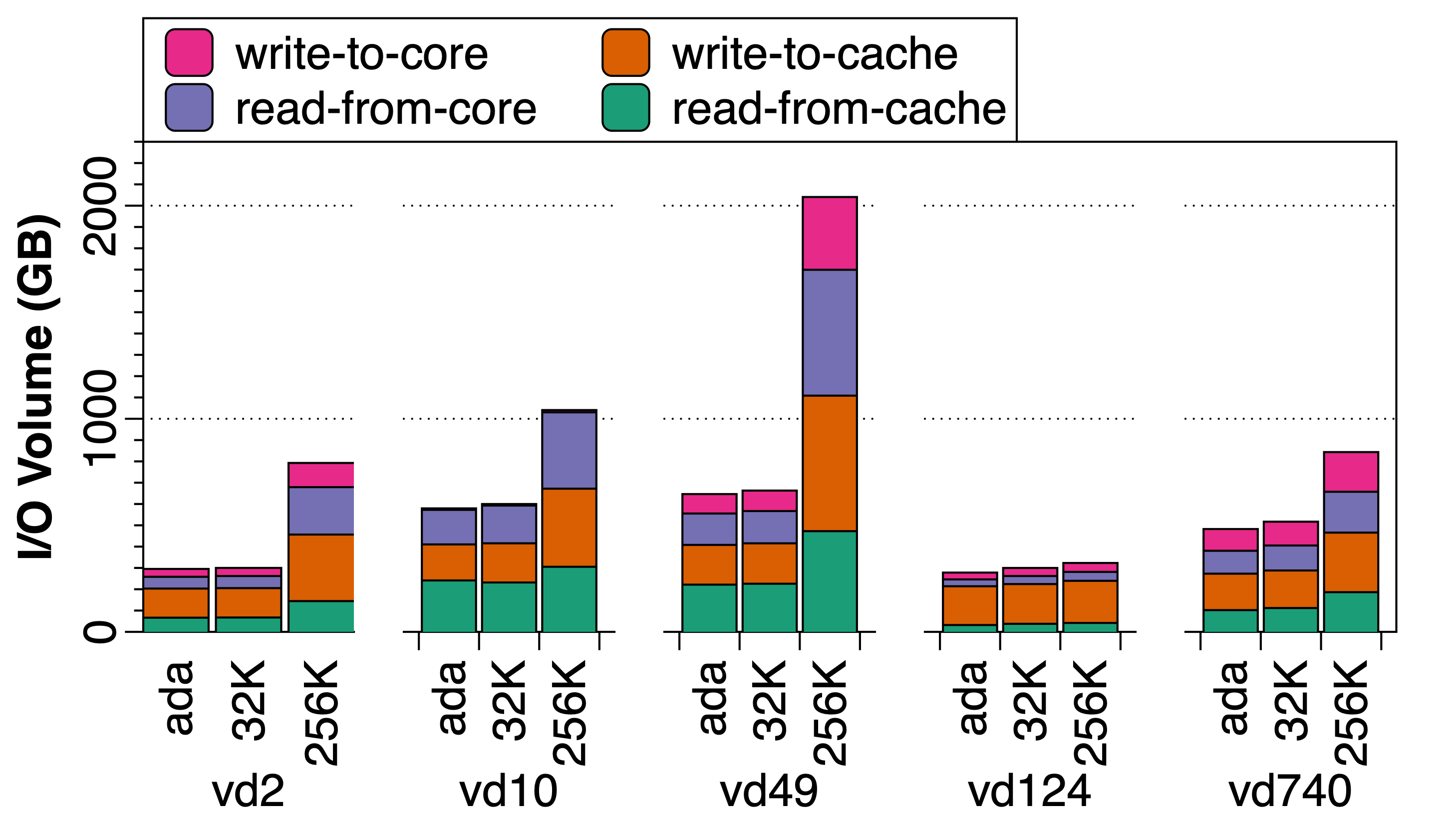}
	 \caption{Alibaba Trace Replay}
	 \label{fig:alibabaiv}
\end{subfigure}
   \hfill
 \begin{subfigure}[b]{0.5\textwidth}
    \includegraphics[width=0.9\textwidth]{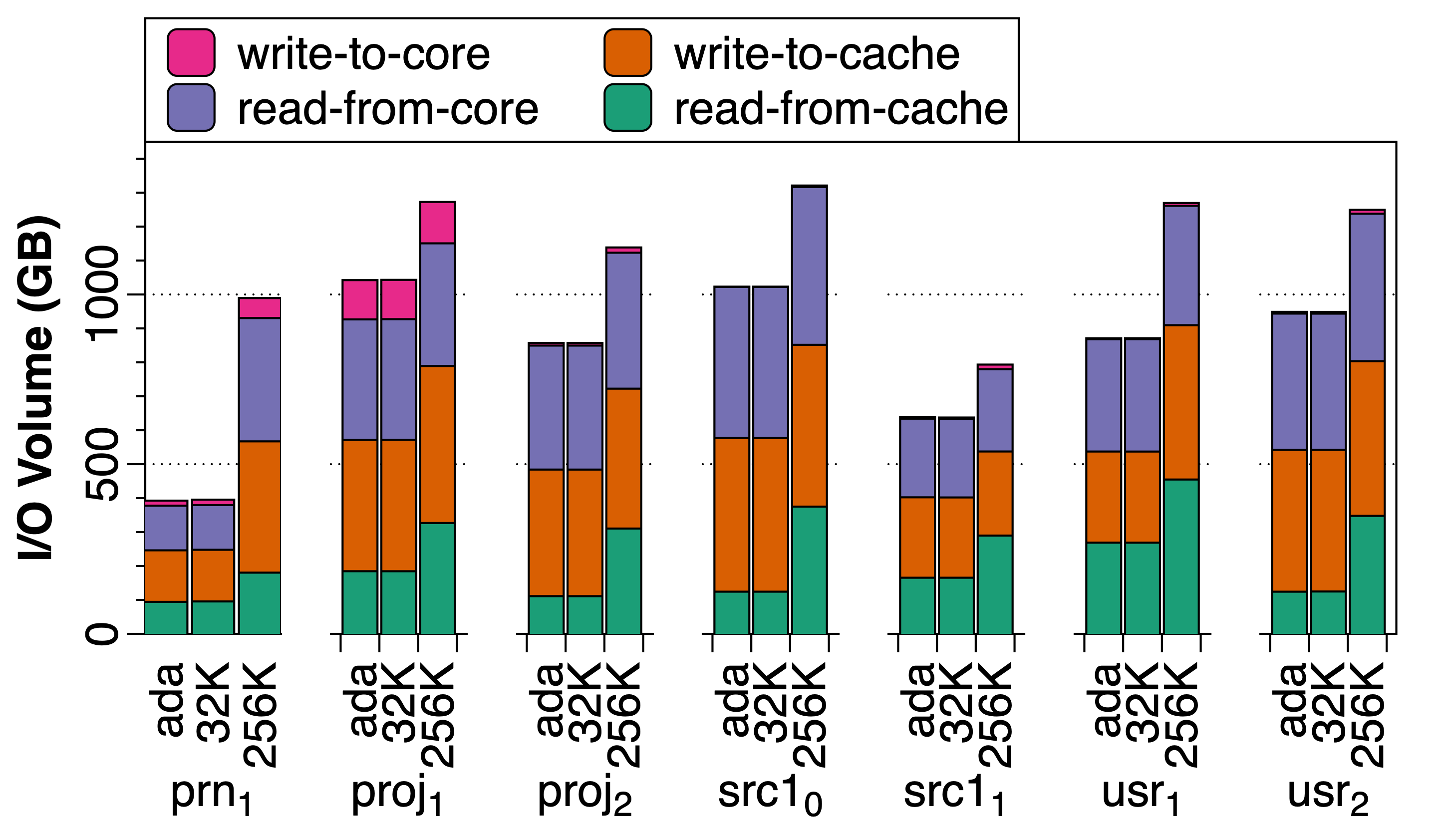}
    \caption{Msr Trace Replay}
    \label{fig:msriv}
 \end{subfigure}
\caption{I/O volumes}
\label{fig:fig4}
\vspace{-10pt}
\end{figure}

\textbf{Average Request Processing Latency.}
Figure~\ref{fig:fig3} shows the average request processing latency from trace replay. This latency is captured from when an I/O request is received by the cache to when a processed I/O request is sent to the storage devices. It includes the latency for the cache block allocation as described in Section~\ref{sec:fix-alloc} and~\ref{sec:var-alloc}. This illustrates the cache block allocation overhead of AdaCache compared to fix-sized caches. Figure~\ref{fig:alibabapl} shows the request processing latency from \textit{alibaba} trace replay. For fix-sized caches, large cache blocks can reduce the number of cache block allocations and therefore reduce the request processing latency. We also observe that AdaCache outperforms 32KiB cache in request processing latency by 25\% for vd124. There are two reasons behind this. First, AdaCache uses large cache blocks for large requests which can help reduce the average request processing latency. Second, the high hit ratio for \textit{alibaba} trace segment (around 70\% for read and 90\% for write) has amortized the extra overhead of adaptive cache block allocation. 

Figure~\ref{fig:systorpl} shows the results from \textit{systor} trace replay. We observe that AdaCache has larger average request process latency than fix-sized caches by 29\% compared to 32KiB cache for LUN1. \textit{Systor} trace segment has around 60\% read hit ratio and because it is read dominant, the low hit ratio fails to amortize the overhead. Although AdaCache brings extra process overhead from adaptive cache block allocation, the overhead is merely a few microseconds and does not hurt the I/O performance as we have seen previously from the I/O latency results. 

\begin{figure}[t]
     \begin{subfigure}[b]{0.5\textwidth}
	\includegraphics[width=0.9\textwidth]{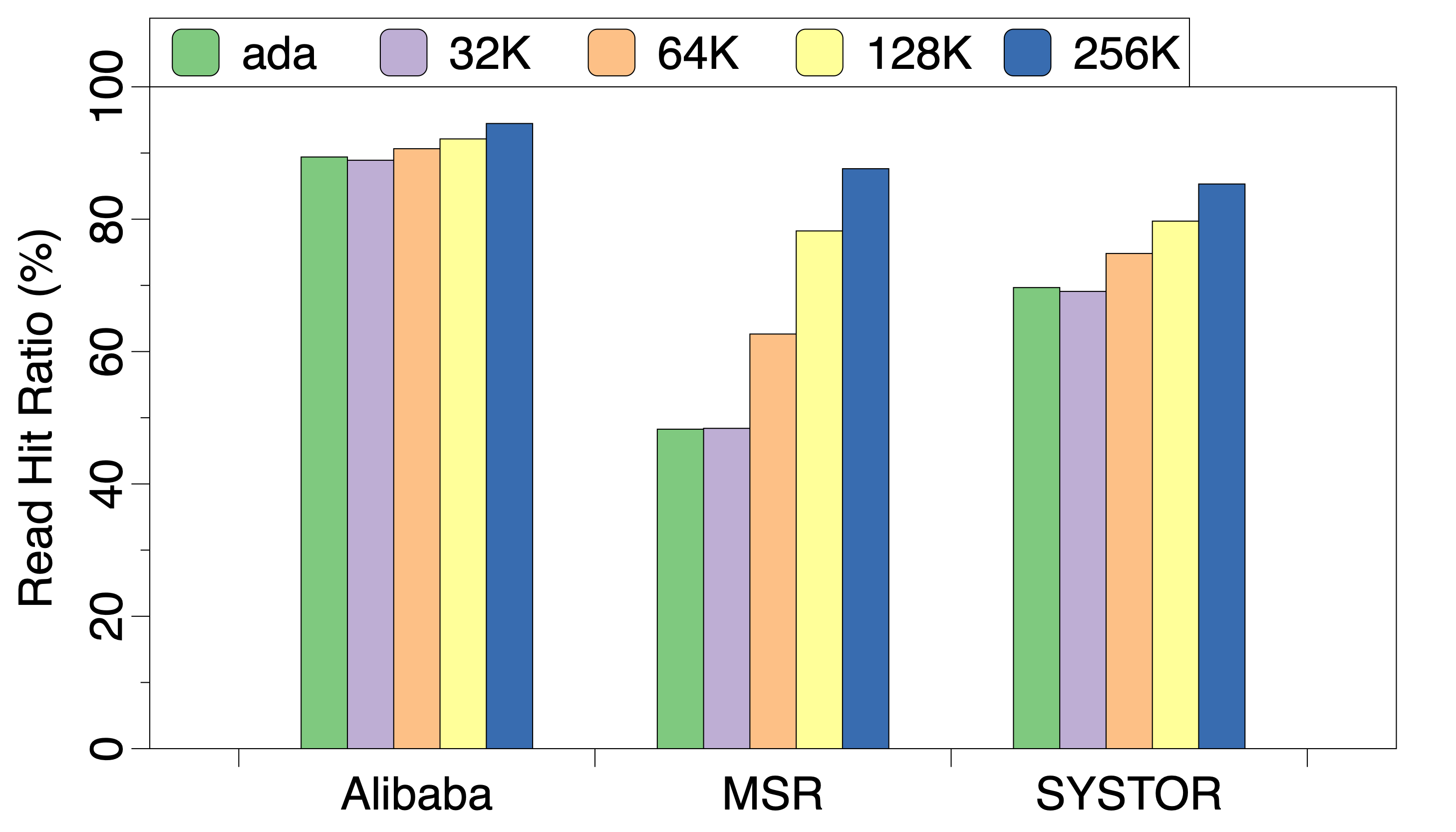}
	 \caption{Read Hit Ratio}
	 \label{fig:arh}
\end{subfigure}
   \hfill
 \begin{subfigure}[b]{0.5\textwidth}
    \includegraphics[width=0.9\textwidth]{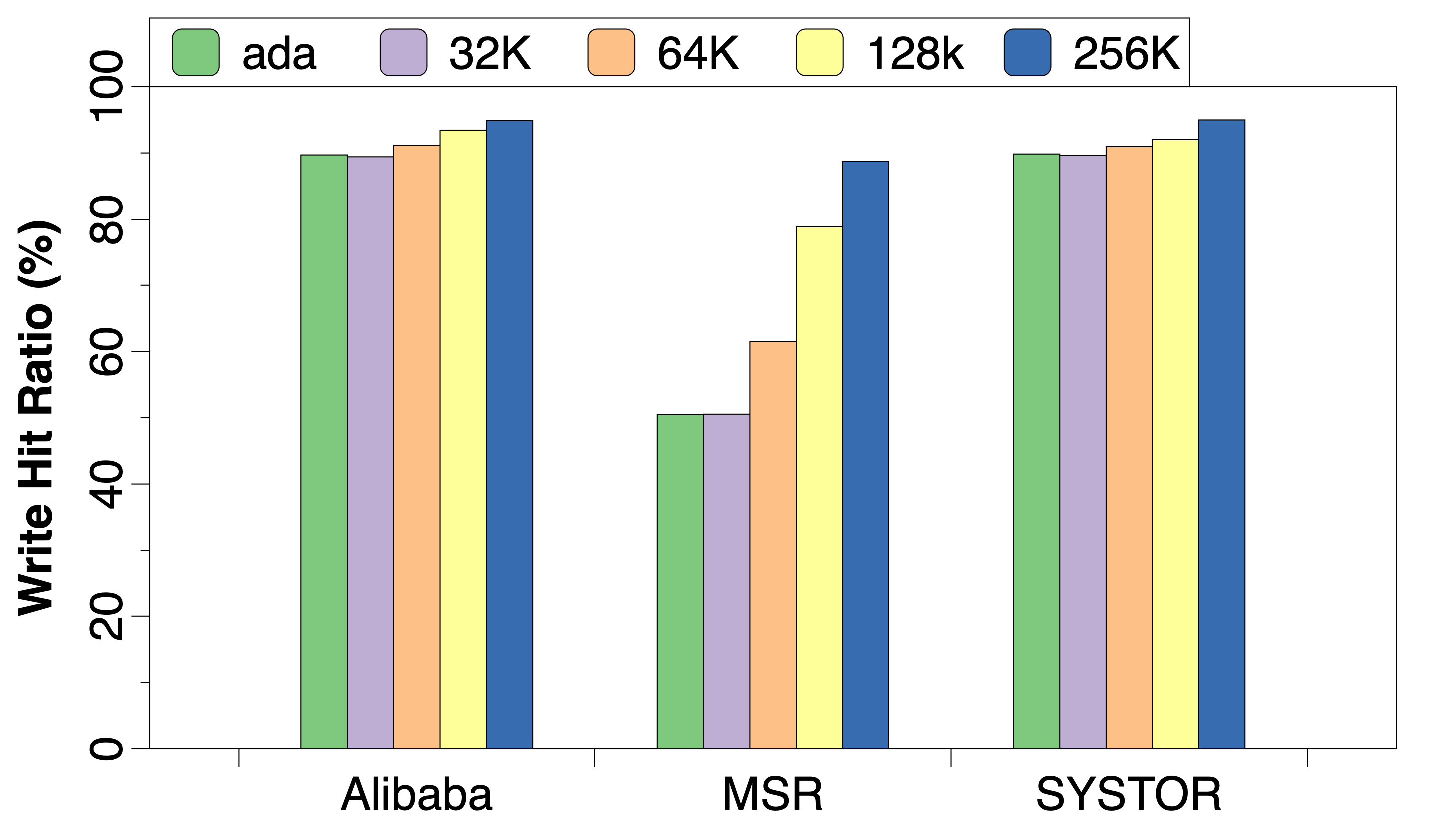}
    \caption{Write Hit Ratio}
    \label{fig:awh}
 \end{subfigure}
\caption{Whole Trace simulation results}
\label{fig:fig5}
\vspace{-10pt}
\end{figure}

\subsection{I/O Volumes}
\label{sec:io}
Figure~\ref{fig:fig4} shows the total I/O volumes from \textit{alibaba} trace replay and \textit{msr} trace replay. The I/O volume consists of writes to the cloud block storage (write-to-core), reads from the cloud block storage (read-from-core), writes to the cache (write-to-cache), and reads from the cache (reads-from-cache). Due to the space limit, we only show 32KiB cache and 256KiB cache I/O volumes which have the smallest and the largest amount of I/O volumes, respectively. As discussed in Section~\ref{sec:background}, using large cache blocks may cache unnecessary data and lead to cache pollution and high cache miss penalty. We also observe that AdaCache has a similar amount of I/O volumes as the 32KiB cache. This is because although it uses large cache blocks, it caches only necessary data based on the request size. It does not suffer from the large cache miss penalty as the 256KiB cache does. Of the four types of I/O volumes, I/Os to cloud block storage has much larger overhead than I/Os to cache. Compared to 256KiB cache, AdaCache can save 74\% I/Os to cloud block storage and 63\% I/Os to cache for vd49 from \textit{alibaba}.

\begin{figure}[t]
	\centering
	\includegraphics[width=0.9\columnwidth]{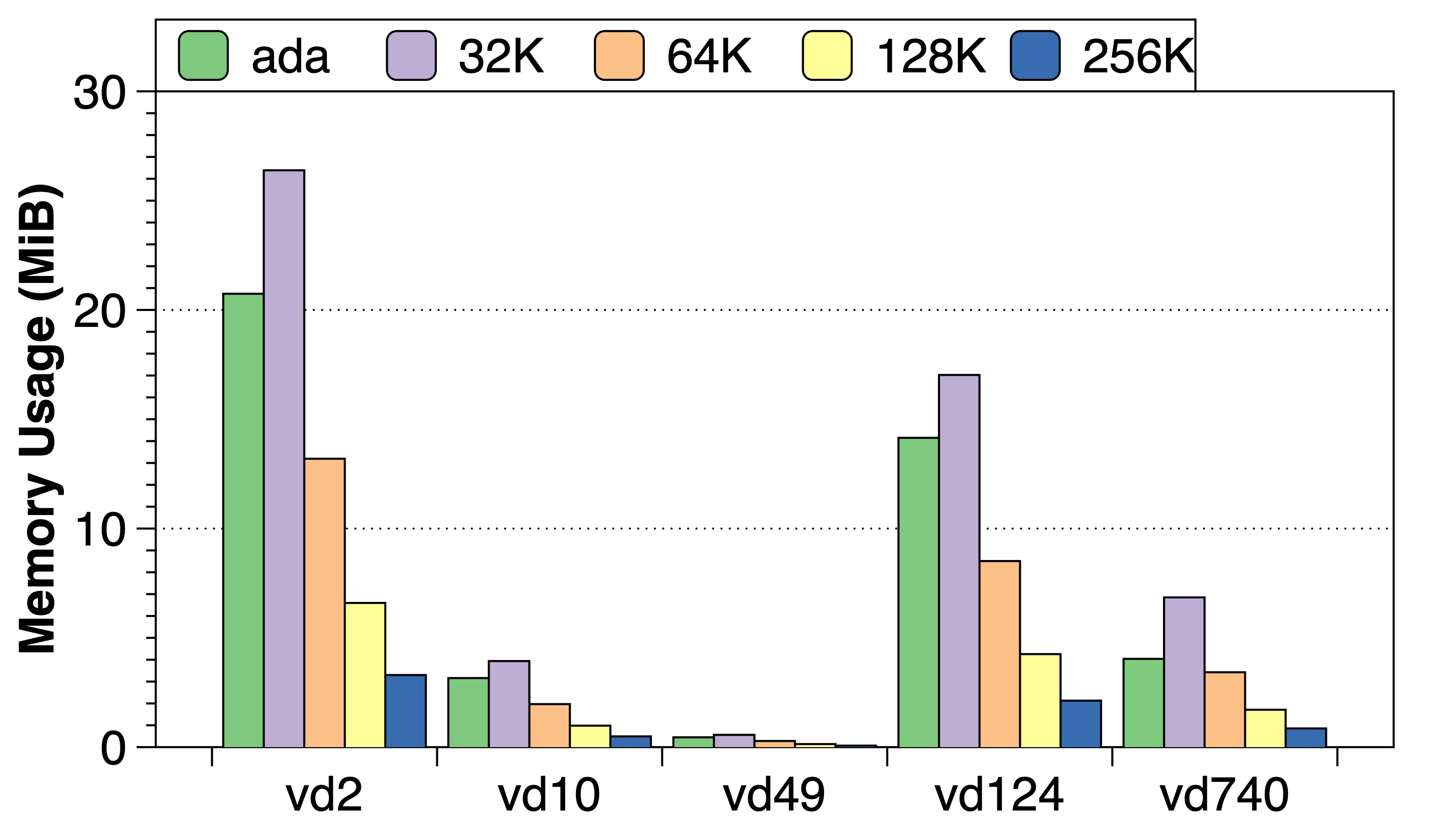}
	\caption{Memory Usage For Alibaba Trace Replay}
	\label{fig:fig6}
 \vspace{-10pt}
\end{figure}

\begin{figure*}[t]
     \begin{subfigure}[b]{0.32\textwidth}
	\includegraphics[width=1.0\textwidth]{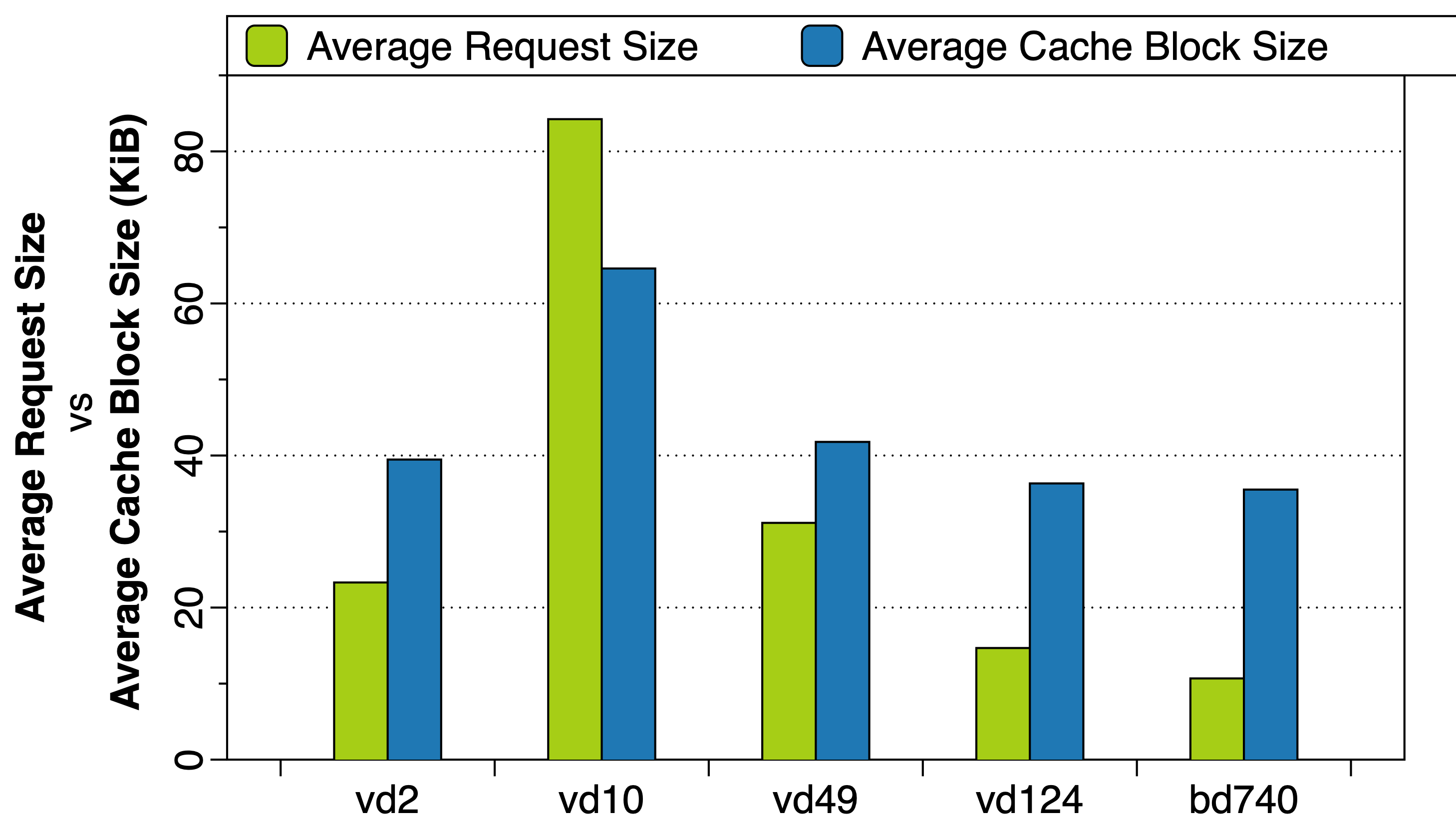}
	 \caption{Alibaba Trace Replay}
	 \label{fig:acbs}
\end{subfigure}
   \hfill
 \begin{subfigure}[b]{0.32\textwidth}
    \includegraphics[width=1.0\textwidth]{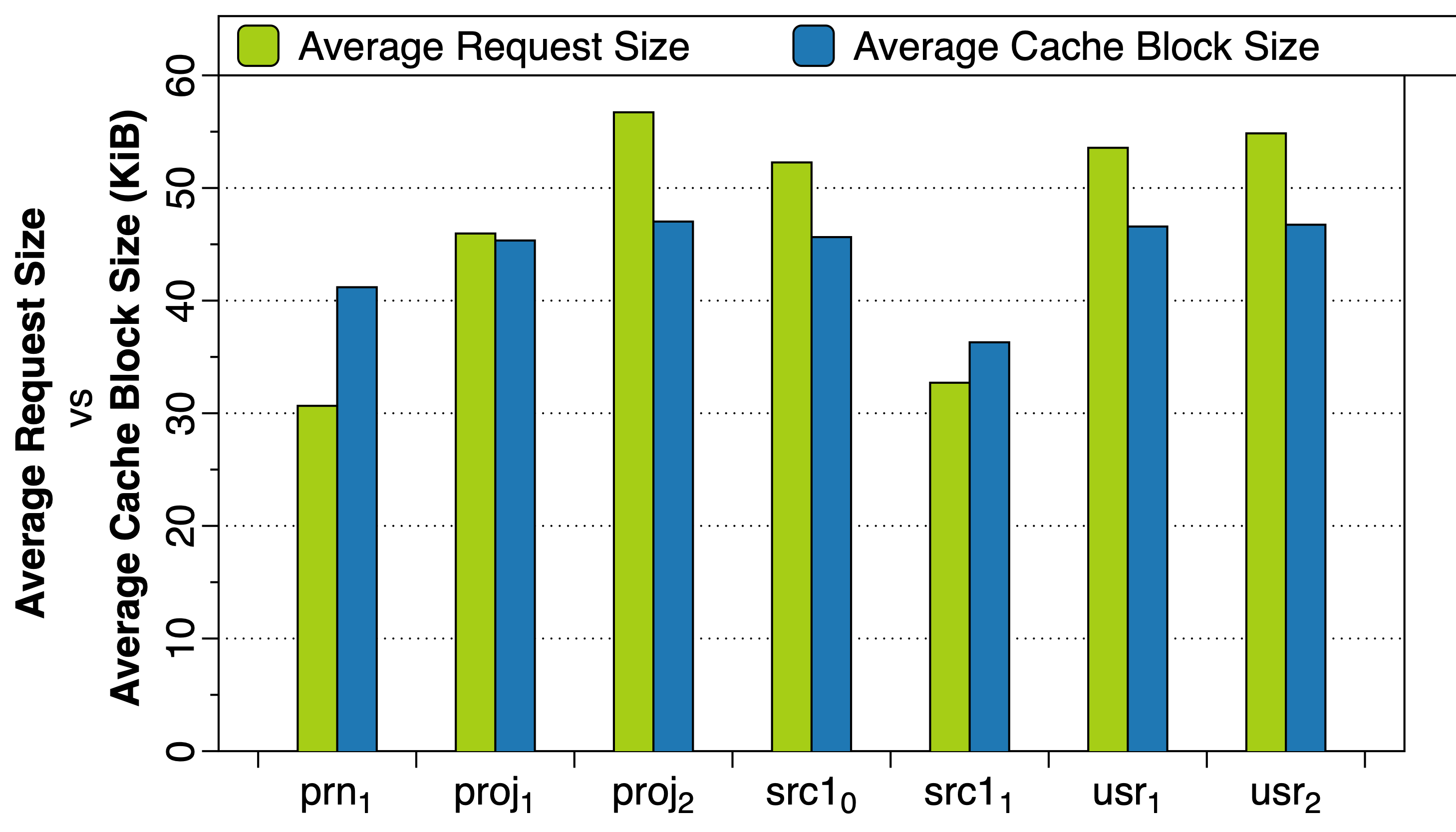}
    \caption{Msr Trace Replay}
    \label{fig:mcbs}
 \end{subfigure}
 \hfill
 \begin{subfigure}[b]{0.32\textwidth}
    \includegraphics[width=1.0\textwidth]{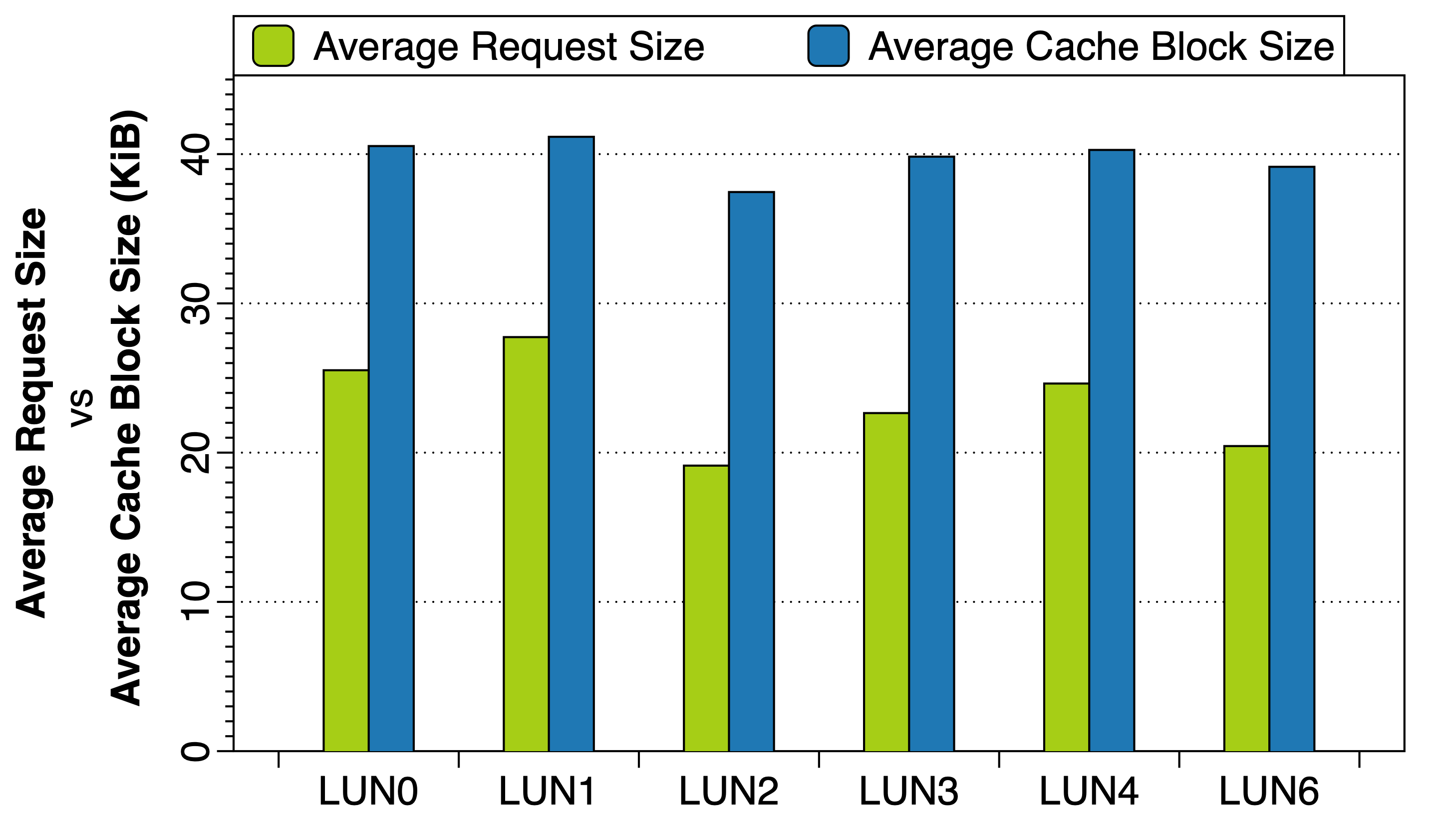}
    \caption{Systor Trace Replay}
    \label{fig:scbs}
 \end{subfigure}
\caption{Average Request Size v.s Average Cache Block Size}
\label{fig:fig7}
\vspace{-10pt}
\end{figure*}

\subsection{Memory Usage}
Figure~\ref{fig:fig6} compares the average metadata memory usage of AdaCache to fix-sized caches during the trace replay of \textit{alibaba}. For larger cache blocks, the number of cache blocks used is smaller which leads to smaller metadata memory usage. AdaCache saves 41\% memory usage compared to 32KiB cache for vd740. From the request size analysis in Section~\ref{sec:background}, \textit{alibaba} mostly consists of small requests. For workloads that have larger requests, AdaCache tends to allocate larger cache blocks and can potentially save more memory.

\subsection{Hit Ratio}
Figure~\ref{fig:fig5} shows the read and write hit ratio from the whole trace simulation of \textit{alibaba}, \textit{msr}, and \textit{systor}. As discussed in Section~\ref{sec:background}, larger cache blocks can benefit from the potential spatial locality within the requests and can achieve better hit ratio compared to smaller cache blocks. We also observe similar behavior when replaying the trace segments. For the whole trace simulation, compared to 256KiB cache, AdaCache has up to 39\% drop in read hit ratio and up to 38\% drop in write hit ratio from \textit{msr}. For trace replay, AdaCache has up to 60\% drop in read hit ratio and up to 59\% drop in write hit ratio from \textit{msr} compared to 256KiB cache. Although the hit ratio is much lower for AdaCache, it has up to 39\% improvement in write performance and 40\% improvement in read performance in trace replay compared to 256KiB cache. This shows that compared to the hit ratio and memory usage, I/O volumes play a more significant role in affecting the cache performance.

\subsection{Effectiveness of Adaptive Cache Block Allocation}
Figure~\ref{fig:fig7} validates the effectiveness of AdaCache block allocation algorithms. It shows two metrics: the average request size for all the missed requests v.s. the average cache block size that AdaCache allocates when a cache miss occurs during trace replay. The core design idea of AdaCache is to adaptively allocate variable-sized cache blocks based on the request size. The differences between these two metrics tell us how well AdaCache follows the design idea. We observe that AdaCache follows the trend of the request size to allocate cache blocks. With larger requests, the average cache block size also gets larger. For small requests which are mostly seen from \textit{alibaba} and \textit{systor}, the average cache block size of AdaCache is bounded by the smallest cache block size 32KiB. For the best case, AdaCache achieves merely a 1\% difference in \textit{msr} trace replay of trace segment proj\_1. 


\vspace{-10pt}
\Section{Related Works}
\vspace{-5pt}
\label{sec:related}
\textbf{Flash Caching.} Flash caching~\cite{scave,centaur,cowcache,jcache} has been extensively studied to improve the I/O performance for slow primary storage systems. Solutions have been proposed to solve the capacity and endurance~\cite{cachededup,nitro}, multi-tenancy~\cite{cloudcache,shards,osca,vcacheshare,jcache} and multi-tier~\cite{karma,unifiedmultilevel,secondtier} problems of flash caching. 
For example, CloudCache~\cite{cloudcache} presents an on-demand cache management solution that meets the performance requirements of each tenant by introducing the Reuse Working Set (RWS) cache demand model. SHARDS~\cite{shards} is an Miss Ratio Curve (MRC) approximation algorithm that focuses on improving MRC efficiency for online cache reassignment by employing uniform randomized spatial sampling. These orthogonal works can be integrated with AdaCache to improve the cache utilization in a disaggregated cloud environment. 
Nitro~\cite{nitro} is a host-side flash cache solution that performs deduplication and compression on the data blocks, after which the compressed variable-sized data chunks are stored in the cache as fixed-size Write-Evict Units (WEUs). Nitro uses LRU at the granularity of WEU for cache replacement. Besides the coarse-grained cache replacement policy employed by both Nitro and AdaCache, AdaCache also uses the fine-grained cache block replacement policy to further improve the cache hit ratio by replacing cold cache blocks inside each group as discussed Section~\ref{sec:design}.

\textbf{Flash Disaggregation.} Storage disaggregation~\cite{ceph, glusterfs, nfs, ebs, minio, s3} is common practice in production environment. High-performance flash disaggregation is also an active research area~\cite{guz2017nvme, nanavati2017decibel}. Since modern NVMe SSDs are significantly faster than SATA SSDs and hard drives, the software overhead becomes nonnegligible. Guz et al.~\cite{guz2017nvme} evaluated the overhead of NVMe SSD storage disaggregation through NVMeoF~\cite{nvmeof} and concluded that the overhead of remote access is negligible compared to local NVMe SSDs. Decibel~\cite{nanavati2017decibel} is a solution for flash storage disaggregation at the rack scale, which follows a design of sharing-nothing and provides virtualized storage with low latency by minimizing the software overhead through the integration of network and storage layers.

\textbf{In-Memory Caching.}
In-memory caching systems~\cite{memcached,redis,aerospike} are widely used in modern software architecture to improve application performance and scalability. For example, Memcached~\cite{memcached} is a lightweight DRAM key-value store that stores key-value pairs of the same value size in slabs of the same slab class. Unlike AdaCache which does global cache block groups replacement, Memcached does time-consuming slab reassignment~\cite{fasterslab,robinhood,lama} across slab classes due to the high concurrency. Data structure optimization~\cite{memc3,fastercuckoo,efficientinmemkv} to save the metadata memory overhead has also been studied. For example, MemC3~\cite{memc3} reduces the metadata memory footprint by up to 30\% for Memcached by using concurrent Cuckoo hashing and CLOCK LRU-approximation cache replacement. These data structure optimization techniques are complementary to AdaCache and can be leveraged to further reduce the metadata memory overhead.

\textbf{Adaptive Cache Block Sizes.} 
The performance impact of varying cache block sizes for both memory and storage cache has been thoroughly studied in literature~\cite{adjustablecoherentcache, linesizechoice, cachesizeperfimpact,10.1145/2155620.2155673,10.5555/52400.52433,10.1145/63404.63407}. However, few have studied the benefits and drawbacks of a cache system with adaptive cache block sizes. Jeremic et al.~\cite{onadapting} proposed a two-size cache block allocation mechanism that employs a small-block and a large-block SSD cache. The source address space is divided into segments of contiguous source blocks where either the small or the large cache block size can be used. The assignment relationship between segments and cache block sizes is adjusted in the background based on the measurement of I/O latency. AdaCache differs from the related work including but not limited to 1) AdaCache supports different numbers of cache block sizes to cater to the workloads' characteristics without delay, 2) AdaCache adapts the cache block size based on the request size which is more efficient and effective than monitoring I/O latency of the system.
To our best knowledge, AdaCache is the first practical storage cache solution using adaptive cache block sizes.


\vspace{-10pt}
\Section{Conclusion}
\vspace{-10pt}
\label{sec:conclusion}
This paper presents a cache system optimized for cloud block storage with constantly changing workloads. The novelties of this work lie in a new cache block allocation design that dynamically adapts the cache block size to the workloads' characteristics. The entire work is cautiously designed to solve the challenges brought by variable-sized cache block allocation.
An extensive experimental evaluation based on real-world block traces confirms that AdaCache can achieve vast improvements in I/O performance and memory usage with negligible run-time overhead. 
{\bibliographystyle{ieeetr}
	\bibliography{main}}

\begin{thebibliography}{10}

\bibitem{ebs}
``Amazon elastic block store (ebs),'' 2023.
\newblock \url{https://aws.amazon.com/ebs/}.

\bibitem{google}
``Persistent disk,'' 2023.
\newblock \url{https://cloud.google.com/persistent-disk}.

\bibitem{ibm}
``Ibm cloud block storage,'' 2023.
\newblock \url{https://www.ibm.com/cloud/block-storage}.

\bibitem{ceph}
``Ceph block device,'' 2023.
\newblock \url{https://docs.ceph.com/en/quincy/rbd/index.html}.

\bibitem{whatis}
``What is block storage,'' 2021.
\newblock \url{https://aws.amazon.com/what-is/block-storage/}.

\bibitem{ebsefss3}
``Ebs pricing and performance: A comparison with amazon efs and amazon s3,''
  2018.
\newblock
  \url{https://bluexp.netapp.com/blog/ebs-efs-amazons3-best-cloud-storage-system}.

\bibitem{9925356}
Q.~Yang, R.~Jin, B.~Davis, D.~Inupakutika, and M.~Zhao, ``Performance
  evaluation on cxl-enabled hybrid memory pool,'' in {\em 2022 IEEE
  International Conference on Networking, Architecture and Storage (NAS)},
  pp.~1--5, 2022.

\bibitem{osca}
Y.~Zhang, P.~Huang, K.~Zhou, H.~Wang, J.~Hu, Y.~Ji, and B.~Cheng, ``Osca: An
  online-model based cache allocation scheme in cloud block storage systems,''
  in {\em Proceedings of the 2020 USENIX Conference on Usenix Annual Technical
  Conference}, pp.~785--798, 2020.

\bibitem{efficientssdcache}
K.~Zhou, Y.~Zhang, P.~Huang, H.~Wang, Y.~Ji, B.~Cheng, and Y.~Liu, ``Efficient
  ssd cache for cloud block storage via leveraging block reuse distances,''
  {\em IEEE Transactions on Parallel and Distributed Systems}, vol.~31, no.~11,
  pp.~2496--2509, 2020.

\bibitem{cloudcache}
D.~Arteaga, I.~Ahmad, J.~Cabrera, S.~Jun, J.~Xu, S.~Xu, S.~Sundararaman,
  M.~Zhao, S.~Zhen, V.~Tarasov, {\em et~al.}, ``Cloudcache: On-demand flash
  cache management for cloud computing,'' in {\em 14th $\{$USENIX$\}$
  Conference on File and Storage Technologies ($\{$FAST$\}$ 16)}, pp.~355--369,
  2016.

\bibitem{cachededup}
W.~Li, G.~Jean-Baptise, J.~Riveros, G.~Narasimhan, T.~Zhang, and M.~Zhao,
  ``Cachededup: In-line deduplication for flash caching,'' in {\em 14th
  $\{$USENIX$\}$ Conference on File and Storage Technologies ($\{$FAST$\}$
  16)}, pp.~301--314, 2016.

\bibitem{understandrackscale}
S.~Legtchenko, H.~Williams, K.~Razavi, A.~Donnelly, R.~Black, A.~Douglas,
  N.~Cheriere, D.~Fryer, K.~Mast, A.~D. Brown, {\em et~al.}, ``Understanding
  rack-scale disaggregated storage.,'' {\em HotStorage}, vol.~17, p.~2, 2017.

\bibitem{nvmeof}
``Nvm express moves into the future,'' 2023.
\newblock
  \url{https://nvmexpress.org/wp-content/uploads/NVMe\_Over\_Fabrics.pdf}.

\bibitem{nvmeofblog}
``Making a case for a disaggregated storage architecture,'' 2021.
\newblock
  \url{https://www.kalrayinc.com/blog/making-case-disaggregated-storage-architecture/}.

\bibitem{10.1145/633625.52433}
S.~Prybylski, M.~Horowitz, and J.~Hennessy, ``Performance tradeoffs in cache
  design,'' {\em SIGARCH Comput. Archit. News}, vol.~16, p.~290–298, may
  1988.

\bibitem{hennessy2011computer}
J.~L. Hennessy and D.~A. Patterson, {\em Computer architecture: a quantitative
  approach}.
\newblock Elsevier, 2011.

\bibitem{spdk}
I.~Corporation, ``{SPDK}: {S}torage {P}erformance {D}evelopment {K}it.''
  \url{https://spdk.io/}, Accessed 2023.

\bibitem{yubometadata}
Y.~Liu, H.~Li, Y.~Lu, Z.~Chen, and M.~Zhao, ``An efficient and flexible
  metadata management layer for local file systems,'' in {\em 2019 IEEE 37th
  International Conference on Computer Design (ICCD)}, pp.~208--216, 2019.

\bibitem{hasfs}
Y.~Liu, H.~Li, Y.~Lu, Z.~Chen, N.~Xiao, and M.~Zhao, ``Hasfs: optimizing file
  system consistency mechanism on nvm-based hybrid storage architecture,'' {\em
  Cluster Computing}, vol.~23, pp.~2501--2515, 2020.

\bibitem{9091319}
K.~Zhou, Y.~Zhang, P.~Huang, H.~Wang, Y.~Ji, B.~Cheng, and Y.~Liu, ``Efficient
  ssd cache for cloud block storage via leveraging block reuse distances,''
  {\em IEEE Transactions on Parallel and Distributed Systems}, vol.~31, no.~11,
  pp.~2496--2509, 2020.

\bibitem{afzal2019load}
S.~Afzal and G.~Kavitha, ``Load balancing in cloud computing--a hierarchical
  taxonomical classification,'' {\em Journal of Cloud Computing}, vol.~8,
  no.~1, p.~22, 2019.

\bibitem{nvmeof10us}
``Nvm express moves into the future,'' 2021.
\newblock
  \url{https://nvmexpress.org/wp-content/uploads/NVMe\_Over\_Fabrics.pdf}.

\bibitem{nvmeofperformance}
``Spdk nvme-of rdma (target \& initiator) performance report release 22.09,''
  2023.
\newblock
  \url{https://ci.spdk.io/download/performancereports/SPDK\_rdma\_mlx\_perf\_report\_2209.pdf}.

\bibitem{inspur}
``Samsung’s poseidon v2 e3.x reference system,'' 2021.
\newblock \url{https://www.inspursystems.com/product/open-storage/}.

\bibitem{fio}
``Fio,'' 2021.
\newblock \url{https://github.com/axboe/fio}.

\bibitem{shards}
C.~A. Waldspurger, N.~Park, A.~T. Garthwaite, and I.~Ahmad, ``Efficient mrc
  construction with shards.,'' in {\em FAST}, vol.~15, pp.~95--110, 2015.

\bibitem{qcache}
J.~Fu, D.~Arteaga, and M.~Zhao, ``Locality-driven mrc construction and cache
  allocation,'' in {\em Proceedings of the 27th International Symposium on
  High-Performance Parallel and Distributed Computing}, pp.~19--20, 2018.

\bibitem{memcached}
``Memcached,'' 2023.
\newblock \url{https://memcached.org}.

\bibitem{memdensity}
``The density, cost, and marketing of semiconductor memory,'' 2021.
\newblock
  \url{https://news.skhynix.com/the-density-cost-and-marketing-of-semiconductor-memory/}.

\bibitem{li2020depth}
J.~Li, Q.~Wang, P.~P. Lee, and C.~Shi, ``An in-depth analysis of cloud block
  storage workloads in large-scale production,'' in {\em 2020 IEEE
  International Symposium on Workload Characterization (IISWC)}, pp.~37--47,
  IEEE, 2020.

\bibitem{narayanan2008write}
D.~Narayanan, A.~Donnelly, and A.~Rowstron, ``Write off-loading: Practical
  power management for enterprise storage,'' {\em ACM Transactions on Storage
  (TOS)}, vol.~4, no.~3, pp.~1--23, 2008.

\bibitem{lee2017understanding}
C.~Lee, T.~Kumano, T.~Matsuki, H.~Endo, N.~Fukumoto, and M.~Sugawara,
  ``Understanding storage traffic characteristics on enterprise virtual desktop
  infrastructure,'' in {\em Proceedings of the 10th ACM International Systems
  and Storage Conference}, pp.~1--11, 2017.

\bibitem{poeidon}
``Poseidonos,'' 2023.
\newblock \url{https://github.com/poseidonos/poseidonos}.

\bibitem{bdev}
``Block device user guide.''
\newblock \url{https://spdk.io/doc/bdev.html}.

\bibitem{glib}
{The GNOME Project}, ``{GLib -- C Utility Library}.''
  \url{https://developer.gnome.org/glib/}, 2023.

\bibitem{slab}
J.~Bonwick {\em et~al.}, ``The slab allocator: An object-caching kernel memory
  allocator.,'' in {\em USENIX summer}, vol.~16, Boston, MA, USA, 1994.

\bibitem{koller2013write}
R.~Koller, L.~Marmol, R.~Rangaswami, S.~Sundararaman, N.~Talagala, and M.~Zhao,
  ``Write policies for host-side flash caches,'' in {\em Presented as part of
  the 11th $\{$USENIX$\}$ Conference on File and Storage Technologies
  ($\{$FAST$\}$ 13)}, pp.~45--58, 2013.

\bibitem{scave}
T.~Luo, S.~Ma, R.~Lee, X.~Zhang, D.~Liu, and L.~Zhou, ``S-cave: Effective ssd
  caching to improve virtual machine storage performance,'' in {\em Proceedings
  of the 22nd International Conference on Parallel Architectures and
  Compilation Techniques}, pp.~103--112, 2013.

\bibitem{centaur}
R.~Koller, A.~J. Mashtizadeh, and R.~Rangaswami, ``Centaur: Host-side ssd
  caching for storage performance control,'' in {\em 2015 IEEE International
  Conference on Autonomic Computing}, pp.~51--60, IEEE, 2015.

\bibitem{cowcache}
J.~Fu, Y.~Lu, J.~Shu, G.~Liu, and M.~Zhao, ``Cowcache: effective flash caching
  for copy-on-write virtual disks,'' {\em Cluster Computing}, vol.~23,
  pp.~623--639, 2020.

\bibitem{jcache}
J.~Fu, Y.~Liu, and G.~Liu, ``Jcache: Journaling-aware flash caching,'' {\em
  IEEE Access}, vol.~8, pp.~61289--61298, 2020.

\bibitem{nitro}
C.~Li, P.~Shilane, F.~Douglis, H.~Shim, S.~Smaldone, and G.~Wallace, ``Nitro: A
  {Capacity-Optimized} {SSD} cache for primary storage,'' in {\em 2014 USENIX
  Annual Technical Conference (USENIX ATC 14)}, (Philadelphia, PA),
  pp.~501--512, USENIX Association, June 2014.

\bibitem{vcacheshare}
F.~Meng, L.~Zhou, X.~Ma, S.~Uttamchandani, and D.~Liu, ``{vCacheShare}:
  Automated server flash cache space management in a virtualization
  environment,'' in {\em 2014 USENIX Annual Technical Conference (USENIX ATC
  14)}, (Philadelphia, PA), pp.~133--144, USENIX Association, June 2014.

\bibitem{karma}
G.~Yadgar, M.~Factor, and A.~Schuster, ``Karma: {Know-it-All} replacement for a
  multilevel cache,'' in {\em 5th USENIX Conference on File and Storage
  Technologies (FAST 07)}, (San Jose, CA), USENIX Association, Feb. 2007.

\bibitem{unifiedmultilevel}
L.~Ou, X.~He, M.~Kosa, and S.~Scott, ``A unified multiple-level cache for high
  performance storage systems,'' in {\em 13th IEEE International Symposium on
  Modeling, Analysis, and Simulation of Computer and Telecommunication
  Systems}, pp.~143--150, 2005.

\bibitem{secondtier}
X.~Li, A.~Aboulnaga, K.~Salem, A.~Sachedina, and S.~Gao, ``{Second-Tier} cache
  management using write hints,'' in {\em 4th USENIX Conference on File and
  Storage Technologies (FAST 05)}, (San Francisco, CA), USENIX Association,
  Dec. 2005.

\bibitem{glusterfs}
A.~Amar, A.~Raja, and V.~Sundararajan, ``Glusterfs: a scalable network
  filesystem,'' in {\em Proceedings of the 6th USENIX Symposium on Operating
  Systems Design and Implementation}, USENIX Association, 2004.

\bibitem{nfs}
R.~Sandberg, D.~Goldberg, S.~Kleiman, D.~Walsh, and B.~Lyon, ``The design and
  implementation of a distributed file system,'' in {\em Proceedings of the
  1985 Summer USENIX Conference}, USENIX Association, 1985.

\bibitem{minio}
M.~contributors, ``{MinIO} - object storage for the next generation.''
  \url{https://min.io/}, 2021.
\newblock Accessed: March 3, 2023.

\bibitem{s3}
{Amazon Web Services}, ``{Amazon S3} - simple storage service.''
  \url{https://aws.amazon.com/s3/}, 2021.
\newblock Accessed: March 3, 2023.

\bibitem{guz2017nvme}
Z.~Guz, H.~Li, A.~Shayesteh, and V.~Balakrishnan, ``Nvme-over-fabrics
  performance characterization and the path to low-overhead flash
  disaggregation,'' in {\em Proceedings of the 10th ACM International Systems
  and Storage Conference}, pp.~1--9, 2017.

\bibitem{nanavati2017decibel}
M.~Nanavati, J.~Wires, and A.~Warfield, ``Decibel: Isolation and sharing in
  disaggregated rack-scale storage.,'' in {\em NSDI}, vol.~17, pp.~17--33,
  2017.

\bibitem{redis}
S.~Sanfilippo, ``Redis.'' \url{https://redis.io/}, 2009.
\newblock Accessed: March 4, 2023.

\bibitem{aerospike}
B.~Bulkowski and S.~Srinivasan, ``Aerospike: Architecture of a real-time
  operational dbms,'' {\em IEEE Data Eng. Bull.}, vol.~36, no.~1, pp.~3--9,
  2013.

\bibitem{fasterslab}
D.~Byrne, N.~Onder, and Z.~Wang, ``Faster slab reassignment in memcached,'' in
  {\em Proceedings of the International Symposium on Memory Systems},
  pp.~353--362, 2019.

\bibitem{robinhood}
D.~S. Berger, B.~Berg, T.~Zhu, S.~Sen, and M.~Harchol-Balter, ``{RobinHood}:
  Tail latency aware caching -- dynamic reallocation from {Cache-Rich} to
  {Cache-Poor},'' in {\em 13th USENIX Symposium on Operating Systems Design and
  Implementation (OSDI 18)}, (Carlsbad, CA), pp.~195--212, USENIX Association,
  Oct. 2018.

\bibitem{lama}
X.~Hu, X.~Wang, Y.~Li, L.~Zhou, Y.~Luo, C.~Ding, S.~Jiang, and Z.~Wang,
  ``{LAMA}: Optimized locality-aware memory allocation for key-value cache,''
  in {\em 2015 USENIX Annual Technical Conference (USENIX ATC 15)}, (Santa
  Clara, CA), pp.~57--69, USENIX Association, July 2015.

\bibitem{memc3}
B.~Fan, D.~G. Andersen, and M.~Kaminsky, ``{MemC3}: Compact and concurrent
  {MemCache} with dumber caching and smarter hashing,'' in {\em 10th USENIX
  Symposium on Networked Systems Design and Implementation (NSDI 13)},
  (Lombard, IL), pp.~371--384, USENIX Association, Apr. 2013.

\bibitem{fastercuckoo}
X.~Li, D.~G. Andersen, M.~Kaminsky, and M.~J. Freedman, ``Algorithmic
  improvements for fast concurrent cuckoo hashing,'' in {\em Proceedings of the
  Ninth European Conference on Computer Systems}, EuroSys '14, (New York, NY,
  USA), Association for Computing Machinery, 2014.

\bibitem{efficientinmemkv}
H.~Chen, H.~Zhang, M.~Dong, Z.~Wang, Y.~Xia, H.~Guan, and B.~Zang, ``Efficient
  and available in-memory kv-store with hybrid erasure coding and
  replication,'' {\em ACM Trans. Storage}, vol.~13, sep 2017.

\bibitem{adjustablecoherentcache}
C.~Dubnicki and T.~J. LeBlanc, ``Adjustable block size coherent caches,'' {\em
  SIGARCH Comput. Archit. News}, vol.~20, p.~170–180, apr 1992.

\bibitem{linesizechoice}
A.~J. Smith, ``Line (block) size choice for cpu cache memories,'' {\em IEEE
  Transactions on Computers}, vol.~C-36, no.~9, pp.~1063--1075, 1987.

\bibitem{cachesizeperfimpact}
S.~Przybylski, ``The performance impact of block sizes and fetch strategies,''
  {\em SIGARCH Comput. Archit. News}, vol.~18, p.~160–169, may 1990.

\bibitem{10.1145/2155620.2155673}
G.~H. Loh and M.~D. Hill, ``Efficiently enabling conventional block sizes for
  very large die-stacked dram caches,'' in {\em Proceedings of the 44th Annual
  IEEE/ACM International Symposium on Microarchitecture}, MICRO-44, (New York,
  NY, USA), p.~454–464, Association for Computing Machinery, 2011.

\bibitem{10.5555/52400.52433}
S.~Prybylski, M.~Horowitz, and J.~Hennessy, ``Performance tradeoffs in cache
  design,'' in {\em Proceedings of the 15th Annual International Symposium on
  Computer Architecture}, ISCA '88, (Washington, DC, USA), p.~290–298, IEEE
  Computer Society Press, 1988.

\bibitem{10.1145/63404.63407}
A.~Agarwal, J.~Hennessy, and M.~Horowitz, ``An analytical cache model,'' {\em
  ACM Trans. Comput. Syst.}, vol.~7, p.~184–215, may 1989.

\bibitem{onadapting}
N.~Jeremic, H.~Parzyjegla, and G.~Muhl, ``On adapting the cache block size in
  ssd caches,'' in {\em 2021 IEEE International Conference on Networking,
  Architecture and Storage (NAS)}, pp.~1--8, 2021.

\end{thebibliography}

\end{document}